\documentclass[10pt]{article}
\usepackage{verbatim,amsmath,amssymb,amsthm}
\usepackage[dvips]{graphics}
\usepackage{epsfig}

\newcommand{\ignore}[1]{}

\newcommand {\beq} {\begin{equation}}
\newcommand {\enq} {\end{equation}}
\newcommand {\ra} {\rangle}
\newcommand {\la} {\langle}

\newtheorem{lemm}{Lemma}\newtheorem{theorem}{Theorem}

\newtheorem{claim}{Claim}
\newtheorem{corol}{Corollary}

\newtheorem{deff}{Definition}
\newtheorem{fact}{Fact}

\newcommand{\ket}[1]{\left|#1\right\rangle}
\newcommand{\eqdef}{\stackrel{\rm def}{=}}

\newcommand{\A}{\mathcal{A}}
\newcommand{\set}[1]{{\left\{#1\right\}}}
\newcommand{\B}{\set{0,1}}

\newcommand{\R}{\mathbb{R}}
\newcommand{\Z}{\mathbb{Z}}

\title{ Adiabatic Quantum State Generation and Statistical Zero Knowledge}
\author{Dorit Aharonov\thanks{Department of Computer Science and Engineering,
Hebrew University, Jerusalem, Israel and mathematical Sciences Research Institute, Berkeley, California, e-mail:doria@cs.huji.ac.il}
\and
Amnon Ta-Shma
\thanks{Department of Computer Science,
Tel Aviv University, Tel-Aviv, Israel, e-mail:amnon@post.tau.ac.il}}
\date{}
\begin{document}

\maketitle
{~}

\large
\begin{abstract}
\large
The design of new quantum algorithms has proven to be an extremely
difficult task. This paper considers a different approach to the problem.
We systematically study 'quantum state generation',
namely, which superpositions can be efficiently generated.
We first show that all problems in Statistical Zero Knowledge (SZK),
a class which contains many languages
that are natural candidates for BQP,
can be reduced to an instance of quantum state generation.
This was known before for graph isomorphism, but we give a general recipe
for all problems in SZK. We demonstrate the reduction from the problem to
its quantum state generation version for three examples:
Discrete log, quadratic residuosity and a gap version of
closest vector in a lattice.

We then develop tools for quantum state
generation. For this task, we define the framework of
'adiabatic quantum state generation' which uses
the language of ground states, spectral gaps and Hamiltonians
instead of the standard unitary gate language.
This language stems from the recently suggested
adiabatic computation model \cite{farhiad} and seems to be
especially tailored for the task of quantum state generation.
After defining the paradigm,
 we provide two basic lemmas for adiabatic quantum state
generation:
\begin{itemize}
\item
The Sparse Hamiltonian lemma, which
gives a general technique for implementing sparse Hamiltonians efficiently, and,
\item
The jagged adiabatic path lemma,  which
gives conditions for a sequence of Hamiltonians to allow efficient
adiabatic state generation.
\end{itemize}

We use our tools to prove
 that any quantum state which can be generated efficiently
in the standard model can also be generated efficiently adiabatically,
and vice versa.
Finally we show how to apply our techniques
to generate superpositions
corresponding to limiting distributions of a large class of Markov chains,
 including the uniform distribution over
 all perfect matchings in a bipartite graph
and the set of all grid points inside high dimensional convex bodies.
These final results draw
 an interesting connection between quantum computation and
rapidly mixing Markov chains.

\end{abstract}

\section{Introduction}
Quantum computation carries the hope of solving in quantum polynomial time
classically intractable tasks.
The most notable success so far is Shor's quantum algorithm for
factoring integers and for finding the discrete log \cite{shor}.
Following Shor's algorithm, several
other algorithms were discovered,
such as Hallgren's algorithm for solving Pell's
equation \cite{hallgren}, Watrous's algorithms for the group black box model
\cite{watrous}, and the Legendre symbol algorithm by Van Dam
 et al \cite{legendre}.
Except for \cite{legendre}, all of these algorithms fall into
the framework of the Hidden subgroup problem, and in fact
use exactly the same quantum circuitry;
The exception, \cite{legendre}, is a different algorithm but also
heavily uses Fourier transforms  and exploits the special algebraic structure
of the problem. Recently, a beautiful new 
algorithm by Childs {\it et. al.}\cite{childs}
was found, which gives an exponential speed up
over classical algorithms using an entirely
 different approach, namely quantum walks. 
The algorithm however, works in the black box model and solves 
a fairly contrived problem. 

 One cannot overstate the importance
of developing qualitatively different quantum algorithmic
 techniques and approaches for the development of the
field of quantum computation.
In this paper we attempt to make a step in that direction
by approaching the issue of quantum algorithms from
a different point of view.

It has been folklore knowledge for a few years already that
the problem of graph isomorphism, which is considered classically hard \cite{gibook}
has an  efficient  quantum algorithm
as long as a certain state, namely the superposition of all graphs
isomorphic to a given graph,
\begin{equation}\label{GI}
|\alpha_G\ra=\sum_{\sigma\in S_n} |\sigma(G)\ra
\end{equation}
can be generated efficiently by a quantum
Turing machine (for simplicity,
we ignore normalizing constants in the above state and in the rest
of the paper). The reason that
generating $|\alpha_G\ra$ suffices is very simple:
For two isomorphic graphs, these states are identical, whereas
for two non isomorphic graphs they are orthogonal.
A simple circuit can distinguish
between the case of orthogonal states and that
of identical states, where the main idea is that if the states are orthogonal
they will prevent the different states of
 a qubit attached to them to interfere.
One is tempted to assume that such a state, $|\alpha_G\ra$,
is easy to construct since the equivalent classical distribution,
namely the uniform distribution over all graphs isomorphic to a certain
graph, can be sampled from efficiently.
Indeed, the state $|\beta_G\ra=\sum_{\sigma\in S_n}
 |\sigma\ra\otimes|\sigma(G)\ra$
can be easily generated by this argument;
However, it is a curious (and disturbing) fact of quantum mechanics that
though $|\beta_G\ra$ is an easy state to generate,
so far no one knows how to generate $|\alpha_G\ra$ efficiently,
 because we cannot {\it forget} the value of  $|\sigma\ra$.

In this paper we systematically study the problem of
quantum state generation. We will mostly be interested
in a restricted version of state generation, namely
generating states corresponding to classical probability distributions,
which we loosely refer to as {\it quantum sampling} (or {\it Qsampling})
from a distribution.
To be more specific, we consider
 {\it the probability distribution of a circuit, $D_C$ },
which is the distribution over the outputs of the classical
circuit $C$ when its inputs are uniformly distributed.
Denote $\ket{C} \eqdef \sum_{z\in \{0,1\}^m} \sqrt{D_C(z)} \ket{z}$.
We define the problem of circuit quantum sampling:
\begin{deff}{\bf Circuit Quantum Sampling (CQS):}

{\bf Input:} ($\epsilon, C$) where $C$ is a description of a
 classical circuit from $n$ to $m$ bits,
and $0 \le \epsilon \le {1 \over 2}$.

{\bf Output:} A description of a quantum circuit $Q$ of size $poly(|C|)$
 such that $|Q(|\vec{0}\ra)-\ket{C}|\le \epsilon$.
\end{deff}

We first show that most of the quantum algorithmic
problems considered so far can be reduced to quantum sampling.
Most problems that were considered good candidates for BQP,
such as discrete log (DLOG),
 quadratic residuosity, approximating closest and shortest
vectors in a lattice, graph isomorphism and more,
belong to the complexity class {\it statistical zero knowledge}, or SZK
(see section \ref{szk} for background.)
We prove

\begin{theorem}
\label{thm:szk}
Any ${\cal L} \in SZK$ (Statistical Zero Knowledge)
 can be reduced to a family of instances
of CQS.
\end{theorem}

The proof relies on a reduction by Sahai and Vadhan \cite{SV97}
from SZK to a complete problem called
statistical difference.
Theorem \ref{thm:szk} shows that a general solution for quantum sampling
would imply $SZK \subseteq BQP$.
We note that there exists an oracle $A$ relative to which  $SZK^A \not \subset BQP^A$ \cite{aaronson},
and so such a proof must be non relativizing.

Theorem \ref{thm:szk}
translates a zero knowledge
proof into an instance of CQS.
In general, the reduction can be quite involved, building on the
 reduction in \cite{SV97}.
Specific examples of special interest turn out to be simpler,
e.g., for the case of graph isomorphism described above,
the reduction results in a circuit
$C_G$ that gets as an input a uniformly random string and outputs
a uniformly random graph isomorphic to $G$.
In section \ref{szk} we demonstrate the reduction
for three interesting cases: a decision variant of DLOG
(based on a zero knowledge proof of Goldreich and Kushilevitz \cite{GK93}),
 quadratic
residuosity (based on a zero knowledge proof of Goldwasser, Micali and Rackoff \cite{GMR89}) and
approximating the closest vector problem in lattices (based on a
zero knowledge proof of Goldreich and Goldwasser \cite{GG98}).
The special cases reveal
that although quite often one can look at the zero knowledge proof
and directly infer the required state generation, sometimes it is
not obvious such a transition exists at all. Theorem
\ref{thm:szk}, however, tells us such a reduction is always
possible.

The problem of what states can be generated efficiently by a quantum computer
is thus of critical importance to the understanding of the
computational power of quantum computers.
We therefore embark on the task of designing tools for
quantum state generation, and studying which states can be
generated efficiently.
The recently suggested framework of
adiabatic quantum computation \cite{farhiad} seems to be tailored
exactly for this purpose, since it is stated in terms of quantum state
generation; Let us first explain this framework.

Recall that the time evolution of a quantum system's
state   $|\psi(t)\ra$ is described by
 Schrodinger's equation: \begin{equation}
i\hbar \frac{d}{dt}|\psi(t)\ra=H(t)|\psi(t)\ra.
\end{equation}
where $H(t)$ is an operator called the {\it Hamiltonian} of the system.
We will consider systems of $n$ qubits;
$H$ is then taken to be {\it local}, i.e.
a sum of operators, each operating on a constant number of qubits.
This captures the physical restriction that interactions in nature
involve only a small number of particles, and means that the Hamiltonian
$H(t)$ can actually be implemented in the lab.
Adiabatic evolution concerns the case in which $H(t)$ varies
very slowly in time;
The qualitative statement of the adiabatic theorem is that if
the quantum system is initialized
in the ground state (the eigenstate with lowest eigenvalue)
of $H(0)$, 
and if the modification of $H$ in time is done slowly enough,
namely   {\it adiabatically},
then the final state will be the ground state of the final
Hamiltonian $H(T)$.

Recently, Farhi, Goldstone, Gutmann and Sipser \cite{farhiad} 
suggested to use adiabatic
evolutions to solve $NP$-hard languages.
It was shown in \cite{farhiad,vandamvaz}
that such adiabatic evolutions can be simulated efficiently
on a quantum circuit,  and so designing such a successful process
would imply a quantum efficient algorithm for the problem.
Farhi {\it et. al.}'s idea was to find
the minimum of a given function $f$ as follows:
$H(0)$ is chosen to be some generic Hamiltonian.
$H(T)$ is chosen to be the {\it problem Hamiltonian}, namely
a matrix which has the values of $f$ on its diagonal and zero everywhere
else.
The system is then initialized in the ground state of $H(0)$
and evolves adiabatically on the convex line
$H(t)=(1-\frac{t}{T})H_0+\frac{t}{T}H_T$.
By the adiabatic theorem if the evolution is slow enough,
the final state will be the groundstate
of $H(T)$ which is exactly the sought after minimum of $f$.

The efficiency of these adiabatic algorithms
is determined by how slow the adiabatic
evolution needs to be for the adiabatic theorem to hold.
It turns out that this
depends mainly on the {\it spectral gaps}
of the Hamiltonians $H(t)$. If these spectral gaps are not too small,
the modification of the Hamiltonians
 can be done 'fairly fast', and the adiabatic algorithm
then becomes efficient.
The main problem in analyzing the efficiency of adiabatic algorithms
is thus lower bounding the spectral gap;
This is a very difficult task in general, and hence not much is known
analytically about adiabatic algorithms.
 \cite{farhi2,farhi3,farhi4} analyze numerically
the performance of adiabatic
algorithms on random instances of NP complete problems.
 It was proven in \cite{vandamvaz, cerf}
 that Grover's quadratic speed up \cite{groversearch} can
 be achieved adiabatically.
Lower bounds for special cases were given in \cite{vandamvaz}.
In \cite{wim} it was shown that adiabatic evolution
with local Hamiltonians is in fact equivalent in computational power
to the standard quantum computation model.

In this paper, we propose to use the language of Adiabatic
evolutions, Hamiltonians,
ground states and spectral gaps as a theoretical
 framework for {\em quantum state
generation}.
Our goal is not to replace the quantum circuit model,
neither to improve on it,
but rather to develop a paradigm, or a language,
in which quantum state generation can be studied conveniently.
The advantage in using the Hamiltonian language is
 that the task of quantum state generation becomes
much more natural, since adiabatic evolution is cast in the language
of state generation.  Furthermore, as we will see,
it seems that this language
lends itself more easily than the standard circuit model
to developing general tools.

In order to provide a framework for the study of state generation 
using the adiabatic language, 
 we define {\it adiabatic quantum state generation}
as general as we can.
Thus, we replace the requirement that the Hamiltonians are on a straight line,
with Hamiltonians on any general path.
Second, we replace the requirement that the Hamiltonians are local,
with the requirement that they are {\it simulatable}, i.e.,
that the unitary matrix $e^{-itH(s)}$ can be approximated
by a quantum circuit to within any polynomial accuracy for any polynomially
bounded time $t$.
Thus, we still use the standard model of quantum circuits in 
our paradigm. However, our goal is to derive quantum circuits solving 
the state generation problem,
from adiabatic state generation algorithms. 
Indeed, any adiabatic state generator can  
be simulated efficiently by a quantum circuit.  We give two proofs 
of this fact. The first proof follows from 
 the adiabatic theorem. The 
second proof is self contained, 
and does not require knowledge of the adiabatic theorem. 
Instead it uses the simple Zeno effect\cite{zeno}, thus
 providing an alternative 
point of view of adiabatic algorithms using measurements 
(Such a path was taken also in \cite{farhimeas}.)
This implies that adiabatic state generators can be used as a 
framework for designing algorithms for quantum state generation.

We next describe two basic and general tools for designing
adiabatic state generators.
The first question that one encounters 
is naturally, what kind of Hamiltonians can be used.
In other words,
when is it possible to simulate, or implement,
 a Hamiltonian efficiently.
To this end we prove the {\it sparse Hamiltonian lemma} which
gives a very general condition for a Hamiltonian
to be simulatable.
A Hamiltonian $H$ on $n$ qubits is row-sparse if the number of non-zero entries
at each row is polynomially bounded.
$H$ is said to be row-computable if there exists a (quantum or classical)
efficient algorithm that given $i$ outputs a list $(j,H_{i,j})$ running over all non zero entries $H_{i,j}$.
As a norm for Hamiltonians we use the spectral norm,
i.e. the operator norm induced by the $l_2$ norm on states.

\begin{lemm}
{\bf (The sparse Hamiltonian lemma)}.
\label{sparse} If $H$ is a row-sparse,  row-computable Hamiltonian on $n$
qubits  and $||H|| \le poly(n)$,
then
$H$ is simulatable.
\end{lemm}

We note that this general lemma
 is useful also in two other contexts: first, in the context
of simulating complicated physical systems on a quantum circuit.
Second, for continuous
quantum walks \cite{walks} which use Hamiltonians.
For example, in \cite{childs} Hamiltonians are used to
derive an exponential quantum speed up using quantum walks.
Our lemma can be used directly to simplify
the Hamiltonian implementation used in \cite{childs} and to remove
the unnecessary constraints (namely
coloring of the nodes) which were assumed for the sake of
 simulating the Hamiltonian.

The next question that one encounters in
designing adiabatic quantum state
generation algorithms concerns bounding the spectral gap,
which as we mentioned before is a difficult task.
We would like to develop tools to
find paths in the Hamiltonian space such that the spectral gaps
are guaranteed to be non negligible,
i.e. larger than $1/poly(n)$.
Our next lemma provides a way to do this
in certain cases.
Denote $\alpha(H)$ to be the ground state of $H$ (if unique.)

\begin{lemm}
{\bf (The Jagged Adiabatic Path lemma)}.
\label{jagged}
Let  $\{H_j\}_{j=1}^{T=poly(n)}$ be a sequence
of simulatable
Hamiltonians on $n$ qubits, all with polynomially bounded norm,
non-negligible spectral gaps
and with groundvalues $0$, such that the inner product between the unique
ground states $\alpha(H_j)$ and $\alpha(H_{j+1})$
is non negligible for all $j$.
 Then there is an efficient quantum algorithm that
takes $\alpha(H_0)$  to within arbitrarily small distance
 from $\alpha(H_{T})$.
\end{lemm}

To prove this lemma, the naive idea is to use the sequence of
Hamiltonians as {\it stepping stones} for the adiabatic
computation, connecting $H_j$ to $H_{j+1}$ by a straight line to
create the path $H(t)$. However this way the spectral gaps along
the way might be small.
Instead we use two simple ideas,  which we can turn into
two more useful tools
for manipulating Hamiltonians for adiabatic state generation.
The first idea is to replace
each Hamiltonian $H_j$ by the Hamiltonian $\Pi_{H_j}$
which is the projection on the
subspace orthogonal to the ground states of $H_j$.
We show how to implement these projections using
 Kitaev's phase estimation
algorithm \cite{kitaevphase}.
  The second useful idea is to connect by straight lines
 projections on states with non negligible inner product.
We show that the Hamiltonians on such a line are guaranteed to
have non negligible spectral gap.
These ideas can be put together to show that
  the jagged adiabatic path connecting the projections $\Pi_{H_j}$
is guaranteed to have sufficiently large spectral gap.

We use the above tools to show that
\begin{theorem}\label{equi}
Any quantum state that can be efficiently generated in the circuit model,
can also be efficiently generated by an adiabatic
state generation algorithm, and vice versa.
\end{theorem}

Thus the question of the complexity
of quantum state generation
is equivalent (up to polynomial terms)
in the circuit model and in the adiabatic state generation model.

In the final part of the paper we demonstrate how our methods
for adiabatic quantum state generation work in
a particularly interesting domain,
namely Qsampling from the limiting distributions of Markov chains.
There is an interesting connection between
rapidly mixing Markov chains and adiabatic computation.
A Markov chain is rapidly mixing if and only if the second eigenvalue gap,
namely the difference between the largest and second largest eigenvalue
of the Markov matrix $M$, is non negligible \cite{mixing}. This clearly bears
resemblance to the adiabatic condition of a non negligible spectral gap,
and suggests to look at Hamiltonians of the form

\begin{equation}
H_M=I-M.
\end{equation}

$H_M$ will be a Hamiltonian  if $M$ is symmetric; if $M$ is not
symmetric but is a reversible Markov chain \cite{lovasz}
 we can still define the Hamiltonian corresponding to it
 (see section \ref{app}.)
The sparse Hamiltonian lemma has as an immediate corollary
that for a special type of Markov chains, which we call
{\it strongly samplable}, the quantum analog of the Markov chain can
be implemented:
\begin{corol}\label{th3}
If $M$ is a strongly samplable Markov chain,
then $H_M$ is simulatable.
\end{corol}
In adiabatic computation one is interested in sequences
of Hamiltonians; We thus consider sequences of strongly samplable
Markov chains.
There is a particularly interesting paradigm in
the study of Markov chains where sequences
of Markov chains are repeatedly used:
Approximate counting \cite{approx}.
In approximate counting the idea is to start from a Markov chain
on a set that is easy to count, and which is contained in a large set
$\Omega$ the size of which we want to estimate; One then slowly
 increases the set on which the Markov chain operates
so as to finally get to the desired set $\Omega$. This paradigm
and modifications of it, in which the Markov chains are modified
slightly until the desired Markov chain is attained, are a
commonly used tool in many algorithms; A notable example is the
recent algorithm for approximating the permanent \cite{perm}. In
the last part of the paper we show how to use our techniques to
translate such approximate counting algorithms in order to quantum
sample from the limiting distributions of the final Markov chain.
We show:

\begin{theorem}\label{approxi} (Loosely:)
Let $A$ be an efficient randomized
algorithm to approximately count a set $\Omega$, possibly
with weights; Suppose $A$
uses slowly varying Markov chains starting from a simple
Markov chain.  Then
there is an efficient quantum algorithm $Q$ that Qsamples
 from the final limiting distribution over $\Omega$.
\end{theorem}

We stress that it is NOT the case that we are interested in a quantum
speed up for
sampling from various distributions but rather we are interested in the
{\it coherent} Qsample of the classical distribution.

We exploit this paradigm to Qsample from
the set of all perfect matchings of a bipartite graph,
using the recent algorithm by Jerrum, Sinclair and Vigoda \cite{perm}.
Using the same ideas we can also Qsample from all linear extensions
of partial orders, using Bubley and Dyer algorithm \cite{bubley},
from all lattice points in
a convex body satisfying certain restrictions
 using Applegate-Kannan technique \cite{applegatekannan}
and from many more states.  We note that some of these states (perhaps all)
can be generated using standard techniques which
exploit the self reducibility of the problem (see \cite{grover}).
We stress however that our techniques are
 qualitatively and significantly
different from previous techniques for generating quantum states,
and in particular do not require
self reducibility.
This can be important for extending this approach to other
quantum states.

In this paper we have set the  grounds for the general study of
the problem of Qsampling and adiabatic quantum state generation,
where we have suggested to use the language of Hamiltonians
and ground states for quantum state generation.
This direction points at very interesting and intriguing
connections between quantum computation and many different areas:
the complexity class SZK and its complete 
problem {\it statistical difference} \cite{SV97},
the notion of adiabatic evolution \cite{kato},
the study of rapidly mixing Markov chains using spectral gaps \cite{lovasz},
quantum walks \cite{childs}, and the study
of ground states and spectral gaps of Hamiltonians in Physics.
 Hopefully, these connections will point at various
 techniques from these areas
which can be borrowed to give more tools for
adiabatic quantum state generation; Notably, the study
of spectral gaps of Hamiltonians in physics is
a lively area with various recently developed techniques
(see \cite{spitzer} and references therein).
It seems that a much deeper understanding of the
adiabatic paradigm is required, in order to solve the most interesting
 open question, namely to design interesting new
quantum algorithms.
An open question which might help in the task is
to present known quantum algorithms, eg. Shor's DLOG algorithm,
or the quadratic residuosity algorithm,
in the language of adiabatic computation, in an insightful way.

The rest of the paper is organized as follows.
We start with the results related to SZK;
We then describe quantum adiabatic computation,
define the adiabatic quantum state generation framework, 
and use the adiabatic theorem to 
 prove that an adiabatic state generator implies a state 
generation algorithm. 
Next we prove our two main tools: 
the sparse Hamiltonian lemma, and the jagged adiabatic
path lemma. We then use these tools to 
 prove that adiabatic state generation is equivalent
to standard quantum state generation. 
Finally we draw the connection to Markov chains and demonstrate
how to use our techniques to Qsample from approximately countable sets.
In the appendix we give the second proof of transforming 
adiabatic state generators to algorithms using the Zeno effect.

\section{Qsampling and SZK}
\label{szk}

We start with some background about Statistical Zero Knowledge.
For an excellent source on this subject, see Vadhan's thesis \cite{V00}
or Sahai and Vadhan \cite{SV97}.

\subsection{SZK}

A pair $\Pi=(\Pi_{Yes},\Pi_{No})$ is a promise problem if
$\Pi_{Yes} \subseteq \set{0,1}^*$, $\Pi_{No} \subseteq \set{0,1}^*$
and $\Pi_{Yes} \cap \Pi_{No} = \emptyset$.
We look at $\Pi_{Yes}$ as the set of all {\em yes} instances,
$\Pi_{No }$ as the set of all {\em no} instances
and we do not care about all other inputs.
If every $x \in \set{0,1}^*$ is in $\Pi_{Yes} \cup \Pi_{No}$
we call $\Pi$ a language.

We say a promise problem $\Pi$ has an interactive proof
with soundness error $\epsilon_s$ and completeness error $\epsilon_c$ if
there exists an interactive protocol between a prover $P$ and a verifier $V$
denoted by $(P,V)$,
where $V$ is a probabilistic polynomial time machine, and

\begin{itemize}
\item
If $x \in \Pi_{Yes}$ 
$V$ accepts with probability at least $1-\epsilon_c$.
\item
If $x \in \Pi_{No}$ then {\em for every} prover $P^*$,
$V$ accepts with probability at most $\epsilon_s$.
\end{itemize}

When an interactive proof system $(\Pi,V)$ for a promise problem $\Pi$ is
run on an input $x$, it produces a distribution over "transcripts"
that contain the conversation between the prover and the verifier.
I.e., each possible transcript appears with some probability (depending on the
random coin tosses of the prover and the verifier).

An interactive proof system $(\Pi,V)$ for a promise problem $\Pi$
is said to be "honest verifier statistical zero knowledge", and in
short HVSZK, if there exists a probabilistic polynomial time
simulator $S$ that for every $x \in \Pi_{Yes}$ produces a
distribution on transcripts that is close (in the $\ell_1$
distance defined below) to the distribution on transcripts in the
real proof. If the simulated distribution is exactly the correct
distribution, we say the proof system is "honest verifier {\em
perfect} zero knowledge, and in short HVPZK.

We stress that the simulator's output is based on the input alone, and the simulator has no access
to the prover.
Also, note that we only require the simulator to produce a
good distribution on inputs in $\Pi_{Yes}$,
and we do not care about other inputs. This is because for
"No" instances there is no correct proof anyway.
We refer the interested reader to
Vadhan's thesis \cite{V00} for rigorous definitions
and a discussion of their subtleties.

The definition of
HVSZK captures exactly the notion of ``zero knowledge'';
If the honest verifier  can simulate the interaction with the prover
by himself, in case the input is in $\Pi$, then he does not learn anything
 from the interaction (except for the knowledge that the input is in $\Pi$).
We denote by HVSZK the class of all promise problems that have an
interactive proof which satisfies these restrictions.
One can wonder whether cheating verifiers can get information from
an honest prover by deviating from the protocol. Indeed, in some
interactive proofs this happens. However,
a general result says that any HVSZK proof can be simulated by
one which does not leak much information
even with dishonest verifiers \cite{GSV98}.
We thus denote by SZK the class of all promise problems which
have interactive proof systems which
are statistically zero knowledge against an
honest (or equivalently a general) verifier.

It is known that $BPP \subseteq SZK \subseteq AM \cap coAM$ and that SZK
is closed under complement. It follows that
SZK does not contain any NP--complete language
unless the polynomial hierarchy collapses.
For this, and other results known about this elegant class, we refer the reader, again,
to Vadhan's thesis \cite{V00}.

\subsection{The complete problem}

Recently, Sahai and Vadhan found a natural complete problem for
the class Statistical Zero Knowledge, denoted by SZK. One nice
thing about the problem is that it does not mention interactive
proofs in any explicit or implicit way. We need some facts about
distances between distributions in order to define the problem.
For two classical distributions $\set{p(x)},\set{q(x)}$ define
their $\ell_1$ distance and their {\it fidelity} (this measure is
known by many other names as well):
\begin{eqnarray*}
|p-q|_1 & = & \sum_x |p(x) - q(x)| \\
F(p,q) & = &  \sum_x \sqrt{p(x) q(x)}
\end{eqnarray*}
We also define the variation distance to be $||p-q|| ={1 \over 2} |p-q|_1$ so that it is
a value between $0$ and $1$.
The following fact is very useful:

\begin{fact}
\label{fact:fidelity-trace} (See \cite{nielsen})
\begin{eqnarray*}
1-F(p,q) \le & ||p-q|| & \le \sqrt{1-F(p,q)^2}
\end{eqnarray*}
or equivalently
\begin{eqnarray*}
1-||p-q|| \le & F(p,q) & \le \sqrt{1-||p-q||^2}
\end{eqnarray*}
\end{fact}

We can now define the complete problem for SZK:

\begin{deff}{\bf Statistical Difference  ($SD_{\alpha,\beta}$)}
\begin{description}
\item [Input]: Two classical circuits $C_0,C_1$ with $m$ Boolean outputs.
\item [Promise]: $||D_{C_0}-D_{C_1}|| \ge \alpha$ or $||D_{C_0}-D_{C_1}|| \le \beta$.
\item [Output]: Which of the two possibilities occurs?
 ({\it yes} for the first case and {\it no} for the second)
\end{description}
\end{deff}

Sahai and Vadhan \cite{SV97,V00} show that for any two constants
$0 \le \beta < \alpha \le 1$ such that even $\alpha^2 > \beta$,
$SD_{\alpha,\beta}$ is complete for SZK \footnote{ Sahai and
Vadhan also show, (\cite{V00}, Proposition 4.7.1) that any promise
problem in HVPZK reduces to $\overline{SD_{1/2,0}}$, where the
line above the class denotes complement, i.e., we swap between the
yes and no instances.}. A well explained exposition can also be
found in \cite{V00}.

\subsection{Reduction from SZK to Qsampling.}
We need a very simple building block.

\begin{claim}
\label{cl:angle}
Let $\psi={1 \over \sqrt{2}} (\ket{0,v}+\ket{1,w})$.
If we apply a Hadamard gate on the first qubit and measure it,
we get the answer $0$
with probability  ${{1 + Real(\la v | w \ra)} \over 2}$
and $1$ with probability ${{1 - Real(\la v | w \ra)} \over 2}$.
\end{claim}

The proof is a direct calculation.
We now proceed to prove Theorem \ref{thm:szk}.

\begin{proof}
Let $C_0,C_1$ be an input to $SD$, $C_0,C_1$ are circuits with $m$ outputs.
It is enough to show that $SD_{1/4,3/4} \in BQP$, given that we can
Qsample from the given circuits.
Let us first assume that we can Qsample from both circuits
 with $\epsilon=0$ error.
We can therefore generate the superposition ${1 \over \sqrt{2}} (\ket{0}\ket{C_0}+\ket{1}\ket{C_1})$.
We then apply a Hadamard gate on the first qubit and measure it.
We use Claim \ref{cl:angle} with
$v=\ket{C_0}$ and $w=\ket{C_1}$. In our case

\begin{equation}
\la v | w \ra=\sum_{z \in \B^m} \sqrt{D_{C_0}(z) D_{C_1}(z)}= F(D_{C_0},D_{C_0})
\end{equation}

We therefore get $0$ with probability ${{1+F(D_{C_0},D_{C_0})} \over 2}$.
Thus,
\begin{itemize}
\item
If $||D_{C_0}-D_{C_1}|| \ge \alpha$, then we measure $0$ with probability
${{1+F(D_{C_0},D_{C_0})} \over 2} \le {1+\sqrt{1-||D_{C_0}-D_{C_1}||^2} \over 2} \le {1+\sqrt{1-\alpha^2} \over 2} $,
while,
\item
If $||D_{C_0}-D_{C_1}|| \le \beta$, then we measure $0$ with probability
${{1+F(D_{C_0},D_{C_0})} \over 2} \ge {2-||D_{C_0}-D_{C_1}|| \over 2} \ge 1-{\beta \over 2}$.
\end{itemize}

Setting $\alpha={3 \over 4}$ and $\beta={1 \over 4}$ we get that
if $||D_{C_0}-D_{C_1}|| \ge \alpha$ we measure $0$ with
probability at most ${1+\sqrt{1-\alpha^2} \over 2} \le 0.831$,
while if $||D_{C_0}-D_{C_1}|| \le \beta$ we measure $0$ with
probability at least $1-{\beta \over 2} \ge {7 \over 8}=0.875$.
Repeating the experiment $O(\log({1 \over \delta}))$ times, we can
decide on the right answer with error probability smaller than
$\delta$. If the quantum sampling circuit has a small error (say
$\epsilon<{1 \over 100}$) then the resulting states are close to
the correct ones and the error introduced can be swallowed by the
gap of the BQP algorithm.
\end{proof}

The above theorem shows that in order to give an efficient quantum
algorithm for any problem in SZK, it is sufficient to find an
efficient quantum sampler from the corresponding circuits.
One can use the theorem to start from a zero knowledge proof for a certain
language, and translate it to a family of circuits which we would
like to Qsample from. Sometimes this reduction can be very easy,
without the need to go through the complicated reduction of Sahai
and Vadhan \cite{SV97}, but in general we do not know that the
specification of the states is easy to derive.
For the
sake of illustration, we give the exact descriptions of the states
required to Qsample from for three examples, in which the reduction
turns out to be much simpler than the general case. These cases are of
particular interest for quantum algorithms: discrete log, quadratic
residuosity and a gap version of Closest vector in a lattice.

\subsection{A promise problem equivalent to Discrete Log}

\begin{description}

\item [The problem]:

Goldreich and Kushilevitz \cite{GK93} define the promise problem $DLP_{c}$ as:

\begin{itemize}
\item Input: A prime $p$, a generator $g$ of $Z_p^*$ and an input
$y \in Z_p^*$.
\item Promise:
The promise is that $x=\log_g(y)$ is in $[1,cp] \cup [{p \over 2}+1,{p \over 2}+cp]$,
\item
Output:
Whether $x \in [1,cp]$ or $x \in [{p \over 2}+1,{p \over 2}+cp]$
\end{itemize}

\end{description}

\cite{GK93} proves that DLOG is reducible to
$DLP_{c}$ for every $0 < c < 1/2$. They also prove that
$DLP_{c}$ has a perfect zero knowledge proof if $0 < c \le {1/6}$.
We take $c=1/6$ and show how to solve $DLP_{1/6}$ with CQS.

\begin{description}
\item [The reduction]:

We assume we can solve the construction problem for the circuit
$C_{y,k}=C_{n,g,y,k}$ that computes $C_{y,k}(i) = y \cdot g^{i}
(\bmod p)$ for $i \in \set{0,1}^k$. The algorithm gets into the
state ${1 \over \sqrt{2}} [~ \ket{0} \ket{C_{g^{p/2+1},\lfloor
\log(p) \rfloor -1}} + \ket{1} \ket{C_{y,\lfloor \log(p) \rfloor
-3}} ~]$ and proceeds as in Claim \ref{cl:angle}.

\item [Correctness]:

We have:

\begin{eqnarray}
\label{eqn:middle}
\ket{ C_{g^{p/2+1},\lfloor \log(p) \rfloor -1} } &
= & {1 \over \sqrt{2^t}} \sum_{i=0}^{2^t-1} \ket{ g^{p/2+i}}
\end{eqnarray}

where $t$ is the largest power of $2$ smaller than $p$. Also, as
$y=g^x$ we have

\begin{eqnarray}
\label{eqn:low}
\ket{ C_{y,\lfloor \log(p) \rfloor -3} } & = & {1 \over
\sqrt{2^{t'}}} \sum_{i=0}^{2^{t'-1}} \ket{g^{x+i}}
\end{eqnarray}

where $t'$ is the largest power of $2$ smaller than $p/8$. Now,
comparing the powers of $g$ in the support of Equations
\ref{eqn:middle} and \ref{eqn:low} we see that

\begin{itemize}
\item
If  $x \in [1,cp]$ then
$\ket{ C_{g^{p/2+1},\lfloor \log(p)
\rfloor -1} }$ and $\ket{ C_{y,\lfloor \log(p) \rfloor -3} }$ have
disjoint supports and therefore
$\la C_{y,\lfloor \log(p) \rfloor -3} |C_{g^{p/2+1},\lfloor \log(p) \rfloor -1}\ra | = 0$, while,
\item
If $x \in [{p \over 2}+1,{p \over 2}+cp]$ then the overlap is large
and $|\la C_{y,\lfloor \log(p) \rfloor -3} | C_{g^{p/2+1},\lfloor \log(p) \rfloor -1}\ra|$
is a constant.
\end{itemize}

\end{description}

\subsection{Quadratic residuosity}

\begin{description}

\item [The problem]:

we denote $x R n$ if  $x=y^2 (\bmod n)$ for some $y$, and $x N n$ otherwise.
The problem QR is to decide on input $x,n$ whether $x R n$.
An efficient algorithm is known for the case of $n$ being a prime,
and the problem is
believed to be hard for $n=pq$ where $p,q$ are chosen at random among
large primes $p$ and $q$.
A basic fact, that follows directly from the Chinese remainder theorem is

\begin{fact}
\label{fact:QR} \

\begin{itemize}
\item
If the prime factorization of $n$ is $n=p_1^{e_1} p_2^{e_2} \ldots p_k^{e_k}$,
then for every $x$
\begin{eqnarray*}
x R n & \Longleftrightarrow & \forall_{1 \le i \le k} ~x R p_i
\end{eqnarray*}
\item
If the prime factorization of $n$ is $n=p_1 p_2 \ldots p_k$ then
every $z \in Z_n$ that has a square root, has the same number of square roots.
\end{itemize}
\end{fact}

We show how to reduce the $n=pq$ case to the CQS
(adopting the zero knowledge proof of \cite{GMR89}).

\item [The reduction]:
We use the circuit $C_a(r)$ that on input $r \in Z_n$ outputs $z=r^2 a ~(\bmod n)$.
Suppose we know how to quantum sample $C_a$ for every $a$.
On input integers $n,x$, $(n,x)=1$, the algorithm
gets into the state
${1 \over \sqrt{2}} [\ket{0} \ket{C_1}+\ket{1}\ket{C_x}]$ and proceeds
as in Claim \ref{cl:angle}.

\item [Correctness]:

We have
\begin{eqnarray}
\ket{C_x} & = & \sum_z \sqrt{p_z} \ket{z}
\end{eqnarray}
where $p_z=\Pr_r (z=r^2 x)$, and

\begin{eqnarray}
\ket{C_1} & = & \sum_{z: zRn} \alpha \ket{z}
\end{eqnarray}
for some fixed $\alpha$ independent of $z$.

\begin{itemize}
\item
If $xRn$ then $z=r^2 x$ is also a square.
Furthermore,  as $(x,n)=1$ we have $p_z=\Pr_r (\mbox{$r$ is a square root of ${z \over x}$})$
and as every square has the same number of square roots,
we conclude that $\ket{C_x}=\ket{C_1}$
and $\la C_x | C_1 \ra =1$.
\item
Suppose $x N n$.
There are only $p+q-1$ integers $r \in Z_n$ that are not co-prime to $n$.
For every $r$ co-prime with $n$, $z=xr^2$ must be a non-residue (or else $x Rn$ as well).
We conclude that $\sum_{z:zRn} p_z \le {p+q \over pq} \approx 0$
and so $\la C_x | C_1 \ra \approx 0$.
\end{itemize}

We note that for a general $n$, different elements might have a different number of solutions
(e.g., try $n=8$) and the number of elements not co-prime to $n$ might be large, so one has
to be more careful.
\end{description}

\subsection{Approximating CVP}
We describe here the reduction to quantum sampling for a gap
problem of CVP (closest vector in a lattice), which builds upon
the statistical zero knowledge proof of Goldreich and Goldwasser
\cite{GG98}. A lattice of dimension $n$ is represented by a basis,
denoted $B$, which
 is an $n \times n$
non-singular matrix over $\R$.
The lattice $\mathcal{L}(B)$ is the set of points $\mathcal{L}(B) = \set{Bc ~|~ c \in \Z^n}$,
i.e., all integer linear combinations of the columns of $B$.
The distance $d(v_1,v_2)$ between two points is the Euclidean distance $\ell_2$.
The distance between a point $v$ and a set $\A$ is $d(v,\A)=\min_{a \in A} d(v,a)$.
We also denote $||S||$ the length of the largest vector of the set $S$.
The closest vector problem, CVP, gets as input an $n$--dimensional lattice $B$
and a target vector $v \in \R^n$. The output should be the point $b \in \mathcal{L}(B)$ closest to $v$.
The problem is NP hard. Furthermore, it is NP hard to approximate the distance
to the closest vector in the lattice to within small factors, and it is easy to approximate
it to within $2^{\epsilon n}$ factor,
 for every $\epsilon>0$. See \cite{GG98} for a discussion.
In \cite{GG98} an (honest prover)
perfect zero knowledge proof for being far away from the lattice is given.
We now describe the promise problem.

\begin{description}

\item [The problem]:

\begin{itemize}
\item
Input: An $n$--dimensional lattice $B$, a vector $v \in \R^n$ and designated distance $d$.
We set $g=g(n)=\sqrt{{n \over c \log n}}$, for some $c>0$.
\item
Promise: Either $d(v,\mathcal{L}(B)) \le d$ or $d(v,\mathcal{L}(B) \ge g \cdot d$.
\item
Output: Which possibility happens.
\end{itemize}

We let $H_{t}$ denote the sphere
of all points in $\R^n$ of distance at most
$t$ from the origin.

\item [The reduction]:
The circuit $C_0$ gets as input a random string, and
outputs the vector $r+\eta$,
where $r$ is a uniformly random point in $H_{2^n ||B \cup \set{v}||} \cap \mathcal{L}(B)$
and $\eta$ is a uniformly random
point $\eta \in H_{{g \over 2} \cdot d}$.
\cite{GG98} explain how to sample such points with almost the right
distribution, i.e. they give a description of an efficient
such  $C_0$.

We remark that the points cannot be randomly chosen from the real
(continuous) vector space, due to precision issues, but
\cite{GG98} show that taking a fine enough discrete approximation
and a large enough cutoff of the lattice  results in an
exponentially small error.
>From now on we work in the continuous world,
bearing in mind that in fact everything is implemented in
a discrete approximation of it.

Now assume we can quantum sample from the circuit $C_0$.
We can then also quantum sample from the circuit $C_v$
which we define to be the same circuit except that the outputs
are shifted by the vector $v$ and become  $r+\eta+v$.
To solve the gap problem the algorithm gets into the state
${1 \over \sqrt{2}}
~[~ \ket{0} \ket{C_0} + \ket{1} \ket{C_1} ~]$
and proceeds as in Claim \ref{cl:angle}.

\item [Correctness]:

If $v$ is far away from the lattice $\mathcal{L}(B)$, then
the calculation at \cite{GG98} shows
that the states $\ket{C_0}$  and $\ket{C_1}$ have no overlap and so
$\la C_0 | C_1 \ra = 0$.

On the other hand, suppose $v$ is close to the lattice,
 $d(v,\mathcal{L}(B)) \le d$.
Notice that the noise $\eta$ has magnitude about $gd$,
and so the spheres around
any lattice point $r$ and around $r+v$ have a large overlap.
Indeed, the argument of \cite{GG98} shows that if we express
 $\ket{C_0}= \sum_z p_z \ket{z}$
and $\ket{C_1}= \sum_z p'_z \ket{z}$ then $|p-p'|_1 \le 1-n^{-2c}$.
We see that $\la C_0 | C_1 \ra = F(p,p') \ge n^{-2c}$.
Iterating the above $poly(n)$ times we get an RQP algorithm,
namely a polynomial quantum algorithm with one sided error.

\end{description}

\section{Physics Background}
\label{background}
This section gives background required for our definition
of adiabatic state generation.
We start with some preliminaries regarding the operator norm
and  the Trotter formula.
We then describe the adiabatic theorem, and
the model of adiabatic computation as defined in \cite{farhiad}.

\subsection{Spectral Norm}
\label{pre}

The operator norm of a linear transformation $T$, induced by the $l_2$ norm
 is called the {\it spectral norm} and is defined by
\begin{eqnarray*}
||T|| & = & \max_{\psi \neq 0} {|T\psi| \over |\psi|}
\end{eqnarray*}
If $T$ is Hermitian or Unitary (in general, if $T$ is normal,
namely commutes with its adjoint)
  than $||T||$ equals the largest absolute value of
its eigenvalues. If $U$ is unitary, $||U|| = 1$.
Also, $||AB|| \le ||A|| \cdot ||B||$.
Finally, if $A=(a_{i,j})$ is a $D \times D$ matrix, then $||A|| \le D^2 ||A||_\infty$
where $||A||_\infty = \max_{i,j} |a_{i,j}|$.

\begin{deff}
We say a linear transformation $T_2$ $\alpha$--approximates a linear transformation
$T_1$ if $||T_1-T_2|| \le \alpha$, and if this happens we write $T_2=T_1+\alpha$.
\end{deff}

\subsection{Trotter Formula}
Say $H=\sum H_m$ with each $H_m$ Hermitian.
Trotter's formula states that one can approximate
$e^{-itH}$ by slowly interleaving executions of $e^{-tH_m}$.
We use the following variant of it:

\begin{lemm}
\cite{nielsen}
Let $H_i$ be Hermitian, $H=\sum_{m=1}^M H_m$.
Further assume $H$ and $H_i$ have bounded norm,  $||H||,||H_i|| \le \Lambda$.
Define
\begin{eqnarray*}
U_{\delta} & = &
[~e^{-\delta i H_1} \cdot e^{-\delta i H_2} \cdot \ldots \cdot e^{-\delta i H_M}~] \cdot
[~e^{-\delta i H_M} \cdot e^{-\delta i H_{M-1}} \cdot \ldots \cdot e^{-\delta i H_1}~]
\end{eqnarray*}
Then
$||U_\delta-e^{-2\delta i H}|| \le O(M \cdot {(\delta \Lambda)}^3)$.
\end{lemm}

Using the $||\cdot||$ properties stated above we conclude:

\begin{corol}
Let $H_i$ be Hermitian, $H=\sum_{m=1}^M H_m$.
Assume $||H||,||H_i|| \le \Lambda$.
Then, for every $t>0$
\begin{eqnarray}
||U_\delta^{t \over 2\delta}-e^{-it H}|| & \le &
O( {t \over 2 \delta} \cdot M \cdot (\delta \Lambda)^3)
\end{eqnarray}
\end{corol}

As $||U_\delta-I|| \le 2M \Lambda \delta$
we also have $||U_\delta^{\lfloor {t \over 2\delta} \rfloor}-U_\delta^{t \over 2 \delta}|| \le 2M \Lambda \delta$
and thus:

\begin{corol}
\label{trotter}
Let $H_i$ be Hermitian, $H=\sum_{m=1}^M H_m$.
Assume $||H||,||H_i|| \le \Lambda$.
Then, for every $t>0$
\begin{eqnarray}
||U_\delta^{\lfloor {t \over 2\delta} \rfloor}-e^{-it H}|| & \le &
O( M \Lambda \cdot \delta + M \Lambda^3 t \cdot \delta^2)
\end{eqnarray}
\end{corol}
 
Notice that for every fixed $t,M$ and $\Lambda$, the error term goes
 down to zero with $\delta$.
In applications, we will pick $\delta$ to be polynomially small,
in such a way that the above error
term is polynomially small.

\subsection{Time Dependent Schrodinger Equation}
A quantum state $|\psi\ra$ of a quantum system evolves in time according to
Schrodinger's equation:
\begin{equation}
i\hbar \frac{d}{dt}|\psi(t)\ra=H(t)|\psi(t)\ra\end{equation}
where $H(t)$ is a Hermitian matrix which is called
the Hamiltonian of the physical system.
The evolution of the state from time $0$ to time $T$ can be described by
integrating Schrodinger's equation over time.
If $H$ is constant and independent of time, one gets
\begin{equation}
|\psi(T)\ra=U(T)|\psi(0)\ra=e^{-iHT}|\psi(0)\ra
\end{equation}
Since $H$ is Hermitian $e^{-iHT}$ is unitary,
and so we get the familiar unitary evolution from quantum circuits.
The time evolution is unitary regardless of whether $H$ is time
dependent or not.

The groundstate of a Hamiltonian $H$ is the eigenstate with the smallest eigenvalue,
and we denote it by $\alpha(H)$.
The spectral gap of a Hamiltonian $H$ is the difference between
the smallest and second to smallest eigenvalues, and we denote it by $\Delta(H)$.

\subsection{The adiabatic Theorem}
In the study of {\it adiabatic evolution}
one is interested in the long time behavior (at large times $T$)
of a quantum system initialized
in the ground state of $H$ at time $0$ when the Hamiltonian of the system,
 $H(t)$ changes very slowly in time, namely {\it adiabatically}.

The qualitative statement of the adiabatic theorem is that
if the quantum system is initialized
in the ground state of $H(0)$, the Hamiltonian at time $0$,
and if the modification of $H$ along the path $H(t)$ in the Hamiltonian space
is done infinitely slowly,
then the final state will be the ground state of the final Hamiltonian $H(T)$.

To make this statement correct, we need to add various conditions
and quantifications.
Historically, the first and simplest adiabatic theorem was found by Born and
Fock in 1928 \cite{born}.
In 1958 Kato \cite{kato} improved the statement to essentially
the statement we use in this paper (which we state shortly), and which is
usually referred to as {\it the adiabatic theorem}.
A proof can be found in \cite{masia}.
For more sophisticated adiabatic theorems see \cite{avron}
and references therein.

To state the adiabatic theorem,
it is convenient and traditional to work with a
re-scaled time $s=\frac{t}{T}$ where $T$ is the total time.
The Schrodinger's equation restated in terms of the re-scaled time
$s$ then reads
\begin{equation}\label{slow}
i \hbar \frac{d}{ds}|\psi(s)\ra=T \cdot H(s)|\psi(s)\ra\end{equation}
where $T=\frac{dt}{ds}$ can be referred to as the
{\it delay schedule}, or the {\it total time}.

\begin{theorem}{\bf (The adiabatic theorem, adapted from
\cite{masia,farhiad})}.
Let $H(\cdot)$ be a function from $[0,1]$ to the vector space of Hamiltonians on $n$ qubits.
Assume $H(\cdot)$ is continuous,
has a unique ground state, for every $s \in [0,1]$,
and is differentiable in all but possibly finitely many points.
Let $\epsilon>0$ and assume that the following adiabatic condition holds for
all points $s \in (0,1)$ in which the derivative is defined:
\begin{equation}\label{cond}
T \epsilon \ge \frac{\|\frac{d}{ds}H(s)\|}{(\Delta(H(s))^2}
\end{equation}

Then, a quantum system that is initialized at time $0$ with the ground state
$\alpha(H(0))$  of $H(0)$, and evolves according to the dynamics of the
Hamiltonians $H(\cdot)$, ends up at re-scaled time $1$ at
a state $|\psi(1)\ra$ that
is within $\epsilon^c$ distance from  $\alpha(H(1))$ for some constant $c>0$.
\end{theorem}

We will refer to equation \ref{cond} as the {\it adiabatic condition}.

The proof of the adiabatic theorem is rather involved.
One way to get intuition about it is by observing how the Schrodinger
equation behaves when eigenstates are considered; If the eigenvalue
is $\lambda$, the eigenstate evolves by a multiplicative factor
$e^{i \lambda t}$, which rotates in time faster the larger the absolute value of
the eigenvalue $\lambda$ is, and so
the ground state rotates the least. The fast rotations
are essentially responsible to
the cancellations of the contributions of the vectors with the
higher eigenvalues, due to interference effects.

\section{Adiabatic Quantum State Generation}
\label{state-generation}
In this section we define our paradigm for quantum {\em state generation},
based on the ideas of adiabatic quantum {\em computation} (and the adiabatic theorem).
We would like to allow as much flexibility as possible
in the building blocks. We therefore allow any Hamiltonian
which can be implemented efficiently by quantum circuits.
We also allow using general Hamiltonian paths
and not necessarily straight lines.
We define:

\begin{deff} {\bf (Simulatable Hamiltonians).}
We say a Hamiltonian $H$ on $n$ qubits is {\it simulatable}
if for every $t > 0$ and every accuracy $0 < \alpha < 1$
the unitary transformation
\begin{equation}
U(t)=e^{-iHt}
\end{equation}
can be approximated to within $\alpha$ accuracy by a
quantum circuit of size $poly(n,t,1/\alpha)$.
\end{deff}

If $H$ is simulatable, then by definition
so is $cH$ for any $0\le c\le poly(n)$.
It therefore follows by Trotter's equation (\ref{trotter}) that
any convex combination of two simulatable, bounded norm
Hamiltonians is simulatable.
Also, If $H$ is simulatable and $U$ is a unitary matrix
that can be efficiently applied by a quantum circuit,
then $UHU^\dagger$ is also simulatable, because
$e^{-itUHU^\dagger}=Ue^{-itH}U^\dagger$.

We note that these rules cannot be applied unboundedly
many times in a recursive way, because the simulation will then blow up.
The interested reader is referred to \cite{nielsen,childs} for a more
complete set of rules
for simulating Hamiltonians.

We now describe an adiabatic path, which is an allowable path in the Hamiltonian space:

\begin{deff} {\bf (Adiabatic path).}
A function $H$ from $s \in [0,1]$ to the vector space of Hamiltonians
on $n$ qubits, is an {\it adiabatic path} if
\begin{itemize}
\item $H(s)$ is continuous,
\item $H(s)$ is differentiable except for finitely many points,
\item $\forall s$ $H(s)$ has a unique groundstate, and
\item $\forall s$, $H(s)$ is simulatable given $s$.
\end{itemize}\end{deff}

The adiabatic theorem tells us that the time evolution
of a system evolving along the adiabatic path will
end with the final ground state, if done slowly enough, namely when the
adiabatic condition holds.
Adiabatic quantum state generation is basically the process of
implementing the Schrodinger's evolution along an adiabatic path,
where we require that the adiabatic condition holds.

\ignore{This slowness of time is captured in our {\it schedule}.

\begin{deff}{\bf (Schedule).}
A {\it schedule} for an adiabatic path is
a monotone function $t:[0,1] \to [0,T]$
such that $t(0)=0$, $t(1)=T$ and
\begin{itemize}
\item  $t(0)=0$, $t(1)=T$, and
\item $t$ is explicit, i.e., can be efficiently computable
(by a classical or a quantum circuit.)
\end{itemize}
\end{deff}}

\begin{deff} {\bf (Adiabatic Quantum State Generation).}
An adiabatic Quantum State Generator $(H_x(s),T,\epsilon)$
 has for every $x \in \set{0,1}^n$
an adiabatic path $H_x(s)$, such that for the given $T,\epsilon$
the adiabatic condition is satisfied for all $s\in[0,1]$ where it is defined.
We also require that the generator is explicit, i.e., that
there exists a polynomial time quantum machine that
\begin{itemize}
\item
On input $x \in \set{0,1}^n$ outputs $\alpha(H_x(0))$, the groundstate of $H_x(0)$, and,
\item
On input
$x \in \set{0,1}^n$, $s \in [0,1]$ and $\delta>0$
outputs \ignore{an approximation of $t_x(s)$, and}
 a circuit $C_x(s)$ approximating $e^{-i \delta H_x(s)}$.
\end{itemize}
We then say the generator adiabatically generates $\alpha(H_x(1))$.
\end{deff}

\noindent{\it Remark:}
We note that in previous papers on adiabatic computation, 
eg. in \cite{vandamvaz}, a delay schedule
$\tau(s)$ which is a function of $s$ was used.
We chose to work with one single delay parameter, $T$, instead,
which might seem restrictive;
However, working with a single parameter does
not restrict the model since more complicated
delay schedules can be encoded into the dependence on $s$.

{~}

We will show that every adiabatic quantum state Generator can be efficiently simulated by a quantum circuit, 
in Claim \ref{cl}.
We later on prove  the other direction of Claim \ref{cl}, which implies
Theorem \ref{equi}, which shows the equivalence in computational power
of quantum state generation in the standard and in the adiabatic
frameworks.
 Thus, designing state generation algorithms in the adiabatic
paradigm indeed makes sense since it can be simulated
efficiently on a quantum circuit, and
we do not lose in computational power by
moving to the adiabatic framework and working only with ground states.
The advantage in working in the adiabatic model is that the language of
this paradigm seems more
adequate for developing general tools.
After the statement and proof of Claim \ref{cl},
we proceed to prove several such basic tools.
Once we develop these tools, we will be able to prove the other direction
of the equivalence
theorem and apply the tools for generating interesting states.

\subsection{Circuit simulation of adiabatic state generation}
An adiabatic state generator can be simulated efficiently by
a quantum circuit:

\begin{claim}\label{cl}
{\bf (Circuit simulation of adiabatic state generation)}. Let
 $(H_x(s),T,\epsilon)$ be an Adiabatic Quantum State Generator.
Assume $T \le poly(n)$. 
Then, there exists a quantum circuit that on input $x$
generates the state $\alpha(H_x(1))$ to within $\epsilon$ accuracy,
with size $poly(T,1/\epsilon,n)$.
\end{claim}

\begin{proof} {\bf (Based on Adiabatic Theorem)}
The circuit is built by discretizing time to  sufficiently
small intervals of length $\delta$,
and then applying the unitary matrices $e^{-iH(\delta j)\delta}$.
Intuitively this should work,
as the adiabatic theorem tells us that a physical
system evolving for time $T$ according to Schrodinger's equation
with the given adiabatic path will end up in a state close to
 $\alpha(H_x(1))$.
The formal error analysis can be done by exactly the same techniques that were used
in \cite{vandamvaz}.
We do not give the details of the proof based on the adiabatic
theorem here, since
the proof of the adiabatic theorem itself is hard to follow.
\end{proof}

We give a second proof of Claim \ref{cl}.
The proof does not require knowledge of the adiabatic theorem. 
Instead, it relies on the Zeno effect\cite{zeno}, and due to its simplicity, 
we can give it in full details. 
We include it in order to give a self contained proof of Claim \ref{cl}, 
and also because we believe 
it gives a different,  illuminating perspective on the adiabatic evolution
from the measurement point of view. 
We note that another approach 
toward the connection between adiabatic computation 
and measurements was taken in \cite{farhimeas}.
The full Zeno based proof appears in Appendix \ref{sec:zeno}. 
Here we give a sketch. 

\begin{proof}{\bf (Based on the Zeno effect)}
As before, we begin at the state $\alpha(H(0))$, and 
the circuit is built by discretizing time to sufficiently
small intervals of length $\delta$. 
At each time step $j$, $j=1,\ldots,R$, instead of simulating 
the Hamiltonian as before we apply a measurement 
determined by $H(s_j)$. Specifically, 
we measure the state in a basis which includes the groundstate $\alpha(H(s_j))$.
If $R$ is sufficiently large, the subsequent Hamiltonians 
are very close in the spectral norm,
 and the adiabatic condition guarantees 
that their groundstates are very close in the Euclidean norm. 
Given that at time step $j$ the state is the groundstate $\alpha(H(s_j))$, 
the next measurement results with very high 
probability in a projection on the new groundstate $\alpha(H(s_{j+1}))$. 
The Zeno effect guarantees that the error probability behaves 
like $1/R^2$, i.e. quadratically in $R$ (and not linearly), 
and so the accumulated error after $R$ steps is still 
small, which implies that the probability that the final state is 
the groundstate of $H(1)$ is very high, if $R$ is taken to be large enough. 
\end{proof}

\section{The Sparse Hamiltonian Lemma}
Our first concern is which Hamiltonians can be simulated efficiently.
We restate the sparse Hamiltonian lemma:

{~}

\noindent {\bf Lemma \ref{sparse}}  {\bf The sparse Hamiltonian lemma}
{\it If $H$ is a row-sparse,  row-computable Hamiltonian on $n$
qubits
and $||H|| \le poly(n)$
then $H$ is simulatable. }

{~}

The main idea of the proof is to explicitly write $H$ as a sum of polynomially many
bounded norm Hamiltonians $H_m$ which are all block diagonal (in a combinatorial sense)
and such that the size of the blocks in each matrix is at most $2 \times 2$.
We then show that each Hamiltonian $H_m$ is simulatable and
use Trotter's formula to simulate $H$.

\subsection{The reduction to $2 \times 2$ combinatorially block diagonal matrices.}

Let us define:

\begin{deff}(Combinatorial block.)
Let $A$ be a matrix with rows $ROWS(A)$ and columns $COLS(A)$.
We say $(R,C) \subseteq ROWS(A) \times COLS(A)$ is a combinatorial block
if $|R|=|C|$, for every $c\in C$, $r\not\in R$, $A(c,r)=0$, and
for every $c\not\in C$, $r\in R$, $A(c,r)=0$.
\end{deff}

$A$ is block diagonal in the combinatorial
sense iff there is some renaming of the nodes under which it becomes block diagonal in the usual sense.
Equivalently,
$A$ is block diagonal in the combinatorial sense iff
there is a decomposition of its rows into
$ROWS(A)=\bigcup_{b=1}^B R_b$, and of its columns $COLS(A) = \bigcup_{b=1}^B C_b$
such that for every $b$, $(R_b,C_b)$ is a combinatorial block.
We say $A$ is $2 \times 2$ combinatorially block diagonal,
if each combinatorial block $(R_b,C_b)$ is at most $2 \times 2$, i.e.,
for every $b$ either $|R_b|=|C_b|=1$ or $|R_b|=|C_b|=2$.

\begin{claim} {\bf (Decomposition lemma)}.
\label{decompose}
Let $H$ be a row-sparse, row-computable Hamiltonian over $n$ qubits,
with at most $D$ non-zero elements in each row.
Then there is a way to decompose $H$ into $H=\sum_{m=1}^{(D+1)^2 n^6} H_m$
where:
\begin{itemize}
\item
Each $H_m$ is a row-sparse, row-computable Hamiltonian over $n$ qubits, and,
\item
Each $H_m$ is $2 \times 2$ combinatorially block diagonal.
\end{itemize}
\end{claim}

\begin{figure}[ht]
\begin{center}
\psfig{file=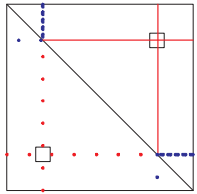,height=1.0in,angle=0}
\caption{\label{fig:2block}\it
In the upper diagonal side of the matrix $H_m$:
the row and column of $(i_b,j_b)$ are empty
because the color $m$ contains the row-index and column index of $(i,j)$,
and the $j_b$'th row and $i_b$'th column are empty because $m$ contains $k, ~i \bmod k,~j \bmod k$
and $i \bmod k \neq j \bmod k$.
The lower diagonal side of $H_m$ is just a reflection of the upper diagonal side.
It follows that $\set{i_b,j_b}$ is a $2 \times 2$ combinatorial block.
}
\end{center}
\end{figure}

\begin{proof}(Of Claim \ref{decompose})
We color all the entries of $H$ with $(D+1)^2 n^6$ colors.
For $(i,j) \in [N] \times [N]$ and $i<j$ (i.e., $(i,j)$ is an upper-diagonal entry)
we define:

\begin{equation}
col_H(i,j)= (k~,~i \bmod k~,~j \bmod k~,~ rindex_H(i,j)~,~cindex_H(i,j))
\end{equation}
where
\begin{itemize}
\item
If $i=j$ we set $k=1$,
otherwise we let $k$ be the first integer in the range $[2..n^2]$
such that $i \neq j (\bmod k)$, and we know there must be such a $k$.
\item
If $H_{i,j}=0$ we set $rindex_H(i,j)=0$,
otherwise we let $rindex_H(i,j)$ be the index of $H_{i,j}$ in the list
of all non-zero elements in the $i$'th row of $H$.
$cindex_H(i,j)$ is similar, but with regard to the columns of $H$.
\end{itemize}
For lower-diagonal entries, $i > j$, we define $col_H(i,j)=col_H(j,i)$.
Altogether, we use $(n^2)^3 \cdot (D+1)^2$ colors.

For a color $m$, we define $H_m[i,j] = H[i,j] \cdot \delta_{col_H(i,j),m}$, i.e., $H_m$ is
$H$ on the entries colored by $m$ and zero everywhere else. Clearly, $H=\sum_m H_m$
and each $H_m$ is Hermitian.
Also as $H$ is row-sparse and row-computable, there is a simple $poly(n)$-time classical algorithm
computing the coloring $col_H(i,j)$, and so each $H_m$ is also row-computable.
It is left to show that it is $2 \times 2$ combinatorially block-diagonal.

Indeed, fix a color $m$.
Let us order all the upper-triangular, non-zero elements of $H_m$ in a list
$NONZERO_m=\set{(i,j) ~|~H_m(i,j) \neq 0~~and~~ i \le j}$.
Say the elements of $NONZERO_m$ are $\set{(i_1,j_1),\ldots,(i_B,j_B)}$.
For every element $(i_b,j_b) \in NONZERO_m$ we introduce a block.
If $i_b=j_b$ then we set $R_b=C_b=\set{i_b}$ while
if $i_b \neq j_b$ then we set $R_b=C_b=\set{i_b,j_b}$.

Say $i_b \neq j_b$ (the $i_b=j_b$ case is similar and simpler).
As the color $m$ contains the row-index and column-index of $(i_b,j_b)$,
it must be the case that $(i_b,j_b)$
is the only element in $NONZERO_m$ from that row (or column).
Furthermore, as $i_b \bmod k \neq j_b \bmod k$,
and both $k,~i \bmod k$ and $j \bmod k$ are included in the color $m$,
it must be the case that there are no elements in $NONZERO_m$
that belong to the $j_b$ row or $i_b$ column (see Figure \ref{fig:2block}).
It follows that $(R_b,C_b)$ is a block.
We therefore see that $H_m$ is $2 \times 2$ combinatorially block-diagonal as desired.
\end{proof}

\begin{claim}
For every $m$, $||H_m|| \le ||H||$.
\end{claim}

\begin{proof}
Fix an $m$. $H_m$ is combinatorially block diagonal
and therefore  its norm $||H_m||$ is achieved as the norm of one of its blocks.
Now, $H_m$ blocks are either
\begin{itemize}
\item
$1 \times 1$, and then the block is $(H_{i,i})$ for some $i$,
and it has norm $|H_{i,i}|$, or,
\item
$2 \times 2$, and then the block is
$\left(
\begin{array}{cc}
0 & A_{k,\ell} \\
A^*_{k,\ell} & 0
\end{array}
\right)$
for some $k,\ell$, and has norm $|A_{k,\ell}|$.
\end{itemize}
It follows that $\max_m ||H_m|| \le \max_{k,\ell} |H_{k,\ell}|$.
On the other hand $||H|| \ge \max_{k,\ell} |H_{k,\ell}|$. The proof follows.
\end{proof}

\subsection{ $2 \times 2$ combinatorially block diagonal matrices are simulatable.}

\begin{claim}
\label{apply}
Every $2 \times 2$ combinatorially block diagonal, row-computable Hamiltonian $A$ is
simulatable to within arbitrary polynomial approximation.
\end{claim}

\begin{proof}
Let $t >0$ and $\alpha>0$ an accuracy parameter.

\begin{description}
\item [The circuit]:

$A$ is $2 \times 2$ combinatorially block diagonal.
It therefore follows that $A$'s action on a given input
$\ket{k}$ is captured by a $2 \times 2$ unitary transformation $U_k$.
Formally, given $k$, say $\ket{k}$ belongs to a $2 \times 2$ block $\set{k,\ell}$ in $A$.
We set $b_k=2$ (for a $2 \times 2$ block)
and $\min_k=\min(k,\ell)$, $\max_k=\max(k,\ell)$ (for the subspace to which $k$ belongs).
We then set $A_k$ to be the part of $A$ relevant to this subspace
$A_k=\left(
\begin{array}{cc}
A_{min_k,min_k} & A_{min_k,max_k} \\
A_{max_k,min_k} & A_{max_k,max_k}
\end{array}
\right)$
and
$U_k=e^{-itA_k}$.
If $\ket{k}$ belongs to a $1 \times 1$ block we similarly define
$b_k=1$, $\min_k=max_k=k$, $A_k=(A_{k,k})$ and $U_k=(e^{-itA_k})$.

Our approximated circuit simulates this behavior.
We need two transformations.
We define
$$T_1 : \ket{k,0} \to \ket{b_k,min_k,max_k,\widetilde{A_k},\widetilde{U_k},k}$$
where $\widetilde{A_k}$ is our approximation to the entries of $A_k$
and  $\widetilde{U_k}$ is our approximation to $e^{-it\widetilde{A_k}}$,
and where both matrices are expressed by their four (or one) entries.
We use $\alpha^{O(1)}$ accuracy.

Having $\widetilde{U_k},min_k,max_k,k$ written down, we can
simulate the action of $\widetilde{U_k}$.
We can therefore have an efficient unitary transformation $T_2$:
$$T_2: \ket{\widetilde{U_k},min_k,max_k} \ket{v}= \ket{\widetilde{U_k},min_k,max_k} \ket{\widetilde{U_k}v}$$
for $\ket{v} \in Span \set{min_k,max_k}$.

Our algorithm is applying $T_1$ followed by $T_2$ and then $T_1^{-1}$ for cleanup.

\item [Correctness]:
Let us denote ${\rm DIFF}= e^{-itA}-T_1^{-1}T_2T_1$.
Our goal is to show that $||{\rm Diff}|| \le \alpha$.
We notice that ${\rm Diff}$ is also $2 \times 2$ block diagonal, and therefore
its norm can be achieved by a vector $\psi$ belonging to one of its dimension one or two subspaces,
say to $Span \set{min_k,max_k}$.
Let $U_k \ket{\psi} = \alpha \ket{min_k}+\beta \ket{max_k}$
and
$\widetilde{U_k} \ket{\psi} = \widetilde{\alpha} \ket{min_k}+\widetilde{\beta} \ket{max_k}$.
We see that:

\begin{eqnarray*}
\ket{\psi,0} & \stackrel{T_1}{\longrightarrow} &
\ket {b_k,min_k,max_k,\widetilde{A_k},\widetilde{U_k},\psi} \\
& \stackrel{T_2}{\longrightarrow} &
\ket {b_k,min_k,max_k,\widetilde{A_k},\widetilde{U_k},\widetilde{U_k} \psi} \\
& = & \widetilde{\alpha} \ket {b_k,min_k,max_k,\widetilde{A_k},\widetilde{U_k},min_k} +
\widetilde{\beta} \ket {b_k,min_k,max_k,\widetilde{A_k},\widetilde{U_k},max_k} \\
& \stackrel{T_1^{-1}}{\longrightarrow} &
\widetilde{\alpha} \ket {min_k,0} +
\widetilde{\beta} \ket {max_k,0}
\end{eqnarray*}
where the first equation holds since it holds for $\ket{min_k}$,
$\ket{max_k}$ and by linearity it holds for the whole subspace spanned by them.
We conclude that $|{\rm Diff} \ket{\psi}| = |(U_k-\widetilde{U_k}) \ket{\psi}|$
and so $||{\rm Diff}||=\max_k ||U_k-\widetilde{U_k}||$.
However, by our construction,
$||\widetilde{A_k}-A_{k}||_\infty \le \alpha^{O(1)}$ and so
$||\widetilde{U_k}-U_{k}|| \le \alpha$ as desired.
\end{description}

\end{proof}

We proved the claim for matrices with $2 \times 2$ combinatorial blocks.
We remark that the same approach works for matrices with $m \times m$ combinatorial blocks,
as long as $m$ is polynomial in $n$.

\subsection{Proving the sparse Hamiltonian lemma}

We now prove the sparse Hamiltonian Lemma:

\begin{proof}
(Of Lemma \ref{sparse}.)
Let $H$ be row-sparse with $D \le poly(n)$ non-zero elements in each row,
and $||H|| =\Lambda \le poly(n)$. Let $t>0$. Our goal is to efficiently simulate $e^{-itH}$
to within $\alpha$ accuracy.

We express $H=\sum_{m=1}^M H_m$ as in Claim \ref{decompose}, $M \le (D+1)^2 n^6 \le poly(n)$.
We choose $\Delta$ such that $O(Mt\Lambda^3\Delta^2) \le {\alpha \over 2}$.
Note that ${1 \over \Delta} \le poly(t,n)$ for some large enough polynomial.
By Claim \ref{apply} we can simulate
in polynomial time each $e^{-i\Delta H_m}$ to within ${\alpha \over 2Mt/\Delta}$
accuracy.
We then compute $U_\Delta^{t \over 2\Delta}$,
using our approximations to $e^{-i \Delta H_m}$, as in Corollary \ref{trotter}.
Corollary \ref{trotter} assures us that our computation is $\alpha$ close to $e^{-itH}$, as desired
(suing the fact that for every $m$, $||H_m|| \le ||H|| = \Lambda \le poly(n)$).
The size of the computation is ${t \over 2\Delta} \cdot 2M \cdot poly(\Delta,M,n,\alpha) = poly(n,t,\alpha)$
as required.
\end{proof}

\section{The Jagged Adiabatic Path Lemma}\label{adiabat}
Next we consider the question of which paths in the Hamiltonian space
guarantee non negligible spectral gaps.
We restate the jagged adiabatic path lemma.

{~}

\noindent{\bf Lemma
\ref{jagged}:}
{\it Let  $\{H_j\}_{j=1}^{T=poly(n)}$ be a sequence
of bounded norm, simulatable
Hamiltonians on $n$ qubits, with non-negligible spectral gaps
and with groundvalues $0$ such that the inner product between the unique
ground states $\alpha(H_j),\alpha(H_{j+1})$
is non negligible for all $j$.
 Then there is an efficient quantum algorithm that
takes $\alpha(H_0)$  to within arbitrarily small distance
 from $\alpha(H_{T})$.}

\begin{figure}[ht]
\begin{center}
\psfig{file=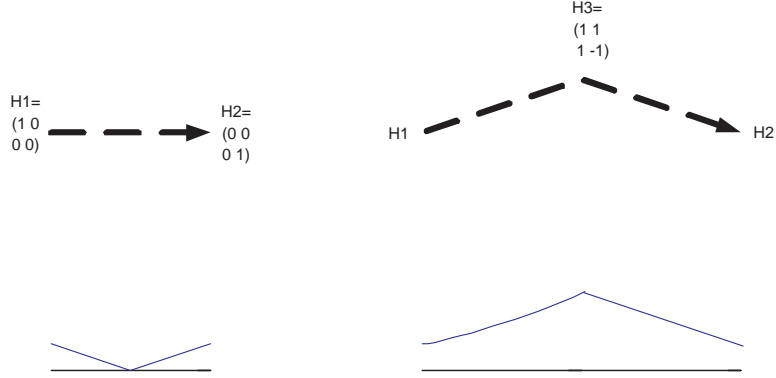,height=2.0in,angle=0}
\caption{\label{fig:jagged}\it
In the left side of the drawing we see two Hamiltonians
$H_1$
and
$H_2$
connected by a straight line, and the spectral gaps
along that line. In the right side of the drawing we see the same two Hamiltonians $H_1$ and $H_2$
connected through a jagged line
that goes through a third connecting Hamiltonian
$H_3$
in the middle, and the spectral gaps along that jagged path.
Note that on the left the spectral
gap becomes zero in the middle, while on the right it is always larger than one.}
\end{center}
\end{figure}

{~}

\begin{proof}(of lemma \ref{jagged})
We replace the sequence $\set{H_j}$ with the sequence
of Hamiltonians
$\set{\Pi_{H_j}}$
where $\Pi_H$ is a projection on
the space orthogonal to the groundstate of $H_j$,
and we connect two neighboring projections by a line.
We prove in claim \ref{proj}, using Kitaev's
phase estimation algorithm, that
the fact that $H_j$ is simulatable implies that so is $\Pi_{H_j}$.
Also, as projections, $\Pi_{H_j}$ have bounded norms, $||\Pi_{H_j}|| \le 1$.
It follows then, by the results mentioned in Section \ref{background},
that all the Hamiltonians on the path connecting these
projections  are simulatable, as convex combinations of simulatable
Hamiltonians.

We now have to show the Hamiltonians on the path have
non negligible spectral gap.
By definition $\Pi_{H_j}$ has a spectral gap equal to $1$.
It remains to show, however, that the Hamiltonians on the line connecting $\Pi_{H_j}$ and $\Pi_{H_{j+1}}$
have large spectral gaps, which we prove in the simple Claim \ref{gap}.

We can now apply the adiabatic theorem and get Lemma \ref{jagged}.
Indeed, a linear time
parameterization suffices to show that this algorithm satisfies
the adiabatic
condition.
\end{proof}

We now turn to the proofs of claims \ref{proj} and \ref{gap}.

\begin{claim}\label{proj}{\bf (Hamiltonian-to-projection lemma)}.
Let $H$ be a Hamiltonian on $n$ qubits such that $e^{-iH}$
 can be approximated to within arbitrary
polynomial accuracy by a polynomial quantum circuit,
and let $\|H\|\le m=poly(n)$.
Let $\Delta(H)$ be non negligible, and larger than $1/n^c$,
and further assume that the groundvalue of $H$ is $0$.
Then the projection $\Pi_H$,
is simulatable.
\end{claim}

\begin{proof}
We observe that Kitaev's phase estimation algorithm
\cite{kitaevphase,nielsen} can be applied here to give a good enough
approximation of the eigenvalue, and as the spectral gap is non-negligible
we can decide
with exponentially good confidence whether an eigenstate
has the lowest eigenvalue or a
larger eigenvalue. We therefore can apply the following algorithm:

\begin{itemize}
\item
Apply Kitaev's phase estimation
algorithm to write down one bit of information on an extra qubit:
whether an input eigenstate of $H$ is the ground state or orthogonal to it.
\item
Apply a phase shift of value $e^{-it}$ to this extra qubit,
conditioned that it is in the state  $|1\ra$ (if it is $|0\ra$ we do nothing).
This conditional phase shift
 corresponds to applying for time $t$ a Hamiltonian
 with two eigenspaces, the ground state and the subspace orthogonal
to it, with respective eigenvalues $0$ and $1$,
which is exactly the desired projection.
\item
Finally, to erase the extra qubit written down, we reverse the first step and
uncalculate the information written on that qubit using Kitaev's
phase estimation algorithm again.
\end{itemize}
\end{proof}

We will also use the following basic but useful claim
regarding the convex combination of two projections.
For a vector $\ket{\alpha}$, the Hamiltonian $H_{\alpha}=I-|\alpha\ra\la\alpha|$
is the projection onto the subspace orthogonal to
$\alpha$. We prove:

\begin{claim}\label{gap}
Let $\ket{\alpha},\ket{\beta}$ be two vectors in some subspace,
$H_{\alpha}=I-|\alpha\ra\la\alpha|$ and
$H_{\beta}=I-|\beta\ra\la\beta|$. For
any convex combination $H_{\eta}=(1-\eta) (I-|\alpha\ra\la\alpha|)+
\eta(I-|\beta\ra\la\beta|, ~~~~\eta\in[0,1]$, of the two Hamiltonians
$H_{\alpha},H_{\beta}$,
$\Delta(H_\eta)\ge |\la \alpha|\beta\ra|$.
\end{claim}

\begin{proof}
To prove this, we observe that
the problem is two dimensional,
write $|\beta\ra=a|\alpha\ra+b|\alpha^\perp\ra$, and
write the matrix $H$ in a basis which contains
$|\alpha\ra$ and $|\alpha^\perp\ra$.
The eigenvalues of this matrix are all $1$ except for a two
dimensional subspace, where the matrix is exactly
\begin{equation}\left(\begin{array}{cc}
\eta|a|^2+(1-\eta) & \eta ab^*\\ \eta a^*b & \eta |b|^2\end{array}\right).\end{equation}
Diagonalizing this matrix we find that the spectral gap is
exactly $\sqrt{1-4(1-\eta)\eta|b|^2}$ which is minimized for $\eta=1/2$
where it is exactly $|a|$.
\end{proof}

We use the tools we have developed to prove the equivalence
of standard and adiabatic state generation complexity,
and for generating interesting
Markov chain states. We start with the equivalence result.

\section{Equivalence of Standard and Adiabatic State Generation}\label{sec:equi}

Theorem \ref{equi} asserts that any quantum state that can be efficiently generated in the quantum circuit model,
can also be efficiently generated by an adiabatic
state generation algorithm, and vice versa.
We already saw the direction from
 adiabatic state generation to quantum circuits.
To complete the proof of Theorem \ref{equi}
we now show the other direction.

\begin{claim}\label{sim}
let $|\phi\ra$ be the final state of a quantum circuit $C$ with $M$
gates, then there is a quantum adiabatic state generator which
outputs this state, of complexity $poly(n,M)$.
\end{claim}

\begin{proof}
W.l.o.g. the circuit starts in the state $\ket{0}$.
We first modify the circuit
so that the state does not change too much between
subsequent time steps.  The reason we need this will become
apparent shortly.
To make this modification,
let us assume for concreteness that the quantum circuit $C$
uses only Hadamard gates, Toffoli gates and Not gates. This
set of gates was recently shown to be universal by Shi \cite{shi},
and a simplified proof can be found in \cite{aharonov2} (Our proof
works with any universal set with obvious modifications.)
We replace each gate $g$ in the circuit by two $\sqrt{g}$ gates.
For $\sqrt{g}$ we can choose any of the possible square roots arbitrarily,
but for concreteness we notice that Hadamard, Not and Toffoli gates have
$\pm 1$ eigenvalues, and
we choose $\sqrt{1}=1$ and $\sqrt{-1}=i$.
We call the modified circuit $C'$.
Obviously $C$ and $C'$ compute the same function.

\begin{description}
\item [The path.]
We let $M'$ be the number of gates in $C'$.
For integer $0 \le j \le M'$, we set
\begin{eqnarray*}
H_x({j \over M'}) & = & I-|\alpha_x(j)\ra\la \alpha_x(j)|
\end{eqnarray*}
where $|\alpha_x(j)\ra$ is the state of the system after applying
the first $j$ gates of $C'$ on the input $x$.
For $s=\frac{j+\eta}{M'}$, $\eta\in [0,1)$, define
$H_x(s)=(1-\eta)H_x(j)+\eta H_x(j+1)$.

\item [The spectral gaps are large.]
Clearly all the Hamiltonians $H_x(j)$
for integer  $0 \le j \le M'$, have non-negligible spectral gaps,
since they are projections.
We claim that for any state $\beta$ and any gate $\sqrt{g}$,
$|\la\beta|\sqrt{g}|\beta\ra| \ge \frac{1}{\sqrt{2}}$.
Indeed, represent $\beta$ as $a_1 v_1+a_2 v_2$
where $v_1$ belongs to the 1-eigenspace of $\sqrt{g}$ and $v_2$ belongs
to the $i$-eigenspace of $\sqrt{g}$. We see that
$|\la\beta|\sqrt{g}|\beta\ra|= ||a_1|^2 +i|a_2|^2|$. As $|a_1|^2+|a_2|^2=1$,
a little algebra shows that
this quantity is at least $\frac{1}{\sqrt{2}}$.
In particular, setting $\beta=\alpha_x(j)$
we see that $|\la \alpha_x(j) | \alpha_x(j+1) \ra| \ge \frac{1}{\sqrt{2}}$.
It therefore follows by claim \ref{gap} that all the Hamiltonians on the line
between $H_x(j)$ and $H_x(j+1)$ have spectral gaps larger than
$\frac{1}{\sqrt{2}}$.

\item [The Hamiltonians are simulatable.]
Given a state $\ket{y}$ we can
\begin{itemize}
\item
Apply the inverse of the first $j$
gates of $C'$,
\item
If we are in state $\ket{x,0}$
apply a phase shift $e^{-i \delta}$, and
\item
Apply the first $j$
gates of $C'$
\end{itemize}
which clearly implements $e^{-i \delta H_x(j)}$.

\item [Adiabatic Condition is Satisfied.]
We have
${dH \over ds}(s_0) =  lim_{\zeta \to 0} {H(s_0+\zeta)-H(s_0) \over \zeta}$.
We ignore the finitely many points $s={j \over M'}$ where $j$ is an {\em integer} in  $[0,M']$.
For all other points $s$, when $\zeta$ goes to $0$ both $H(s_0+\zeta)$ and $H(s_0)$ belong to the same interval.
Say they belong to the $j$'th interval,
$s_0={{j + \eta} \over M'}$, $0<\eta<1$.
Then, 
\begin{eqnarray*}
H(s_0) & = & (1-\eta)H_x(j)+\eta H_x(j+1)  \\
H(s_0+\zeta) = H({{j + \eta + M' \zeta} \over M'}) & = & (1-\eta-M' \zeta)H_x(j)+(\eta+M' \zeta) H_x(j+1)
\end{eqnarray*}
It follows that
$H(s_0+\zeta)-H(s_0) = M' \zeta H_x(j+1) - M' \zeta H_x(j)$
and   ${dH \over ds}(s_0)= M' \cdot [ H_x(j+1) - H_x(j) ]$.
We conclude that $||{dH \over ds}|| \le 2M'$
and to satisfy Equation (\ref{cond}) we just need to pick
$T=O(\frac{M'}{\epsilon})$.
\end{description}
\end{proof}

\section{Quantum State Generation and Markov Chains}\label{app}
Finally, we show how to use our techniques to generate
interesting quantum states related to Markov chains.

\subsection{Markov chain Background}
We will consider Markov chains with states indexed by
$n$ bit strings.  If $M$ is an ergodic (i.e. connected, aperiodic)
Markov chain, characterized with the matrix
 $M$ operating on probability distributions over the state space,
 and $p$ is an initial probability distribution,
 then $\lim_{t\longmapsto\infty}pM^t=\pi$
where $\pi$ is called the limiting distribution and is unique and independent of $p$.

A Markov chain $M$ has eigenvalues between $-1$ and $1$.
A Markov chain is said to be {\it rapidly mixing}
if starting from any initial distribution,
the distribution after polynomially many time steps is within
$\epsilon$ total variation distance from the limiting distribution $\pi$.
\cite{alon} shows that a Markov chain is rapidly mixing if and only if
its second eigenvalue gap is non negligible, namely bounded from below
by $1/poly(n)$.

A Markov chain is {\em reversible} if for the limiting distribution
$\pi$ it holds that
$M[i,j] \cdot \pi_i =M[j,i] \cdot \pi_j$.
We note that a symmetric Markov chain $M$ is in particular reversible.
Also, for an ergodic, reversible Markov chain $M$ $\pi_i>0$ for all $i$.

In approximate counting algorithms
one is interested in sequences of rapidly mixing
Markov chains, where subsequent Markov chains have quite similar
limiting distributions.
For more background regarding Markov chains, see \cite{lovasz}
and references therein;
For more background regarding approximate counting algorithms see
\cite{approx}.

\subsection{Reversible Markov chains and Hamiltonians}
For a reversible $M$ we define
\begin{equation}
H_M= I- Diag(\sqrt{\pi_i})\cdot M \cdot Diag(\frac{1}{\sqrt{\pi_j}})
\end{equation}
A direct calculation shows that $M$ is reversible iff $H_M$ is symmetric.
In such a case we call $H_M$
the {\em Hamiltonian corresponding to $M$}.
The properties of $H_M$ and $M$ are very much related:

\begin{claim}\label{mc}
If $M$ is a reversible Markov chain, we have:
\begin{itemize}
\item $H_M$ is a Hamiltonian with $||H_M || \le 1$.
\item The spectral gap of $H_M$ equals the
second eigenvalue gap of $M$.
\item
If $\pi$ is the limiting distribution of $M$, then the ground state of $H_M$ is
$\alpha(H_M) = \ket{\pi} \eqdef \sum_i \sqrt{\pi(i)} \ket{i}$.
\end{itemize}
\end{claim}

\begin{proof}
If $M$ is reversible, $H_M$ is Hermitian and hence has an eigenvector basis.
In particular $I-H_M= \sqrt{\Pi} M \sqrt{\Pi}^{-1}$ and so
$I-H_M$ and $M$ have the same spectrum. It follows that if the eigenvalues
of $H_M$ are $\set{\lambda_r}$ then the eigenvalues of $M$ are
$\set{1-\lambda_r}$. 
As a reversible Markov chain, $M$ has norm bounded by $1$.

Also, if $v_r$ is an eigenvector of $H_M$ with eigenvalue $\lambda_r$,
then  $Diag(\sqrt{\pi}) v_r$ is the corresponding left eigenvectors of $M$
with eigenvalue $1-\lambda_r$.
In particular, $Diag(\sqrt{\pi}) \alpha(H_{M}) = \pi(M)$.
It therefore follows that $\alpha(H_{M})_i=\sqrt{\pi_i}$, or in short $\alpha(H_{M})=\ket{\pi}$.
\end{proof}

This gives a direct connection between Hamiltonians, spectral gaps and
groundstates on one hand, and rapidly mixing reversible Markov chains
and limiting distribution on the other hand.

\subsection{Simulating $H_M$}
Not every Hamiltonian corresponding to a reversible Markov chain
can be easily simulated.
We will shortly see that the Hamiltonian corresponding to
 a symmetric Markov chain is simulatable. For general
reversible Markov
chains we need some more restrictions. We define:

\begin{deff}
\label{strong}
A reversible Markov chain is
{\it strongly samplable} if it is:
\begin{itemize}
\item
row computable, and,
\item
Given  $i,j\in \Omega$,  there is an efficient way to
approximate ${\pi_i \over \pi_j}$.
\end{itemize}
\end{deff}

Row computability holds in most interesting cases but the second requirement
is quite restrictive. Still, we note that it
 holds in many interesting cases such as all Metropolis algorithms
(see \cite{metro}). It also trivially holds for symmetric $M$, where the
limiting distribution is uniform.

As $H_M [i,j] = \sqrt{\pi_i \over \pi_j} M[i,j]$ we see that if $M$ is strongly samplable
then $H_M$ is row-computable. As $H_M$ has bounded norm, the sparse Hamiltonian lemma implies:

~

\noindent{\bf Corollary \ref{th3}:}
{\it If a Markov chain $M$ is a strongly samplable
then $H_M$ is simulatable.}

\subsection{From Markov chains to Quantum Sampling}

We are interested in strongly samplable rapidly mixing Markov chains,
so that the Hamiltonians are simulatable and have non negligible
spectral gaps by claim \ref{mc}.
To adapt this setting to adiabatic algorithms, and to the setting of the jagged adiabatic path lemma in particular,
we now consider sequences of Markov chains, and define:

\begin{deff} {\bf (Slowly Varying Markov Chains)}.
Let $\{M_t\}_{t=1}^T$ be a sequence of Markov chains on $\Omega$,
$|\Omega|=N=2^n$.
Let $\pi_t$ be the limiting distribution of $M_t$.
We say the sequence is {\em slowly varying}
if  for all $c>0$, for all large enough $n$, for all $1 \le t \le T$
$|\pi_t-\pi_{t+1}|\le 1-1/n^c$.
\end{deff}

We prove that we can move from sequences of slowly varying
Markov chains to Quantum sampling.
We can now state Theorem \ref{approxi} precisely.

{~}

\noindent{\bf Theorem \ref{approxi}: }{\it
Let $\{M_t\}_{t=1}^T$ be a slowly varying sequence of
strongly samplable Markov chains which are all rapidly mixing,
and let $\pi_t$ be their corresponding limiting distributions.
Then if there is an efficient Qsampler for $\ket{\pi_0}$,
then there is an efficient Qsampler for $\ket{\pi_T}$.}

{~}

\begin{proof}
We already saw the Hamiltonians $H_{M_t}$ are simulatable and have bounded norm.
Also, as the Markov chains in the sequence are rapidly mixing, they have large spectral
gaps, and therefore so do the Hamiltonians $H_{M_t}$.
To complete the proof we show that the inner product between the groundstates of subsequent
Hamiltonians is non negligible, and then the theorem follows from
the jagged path lemma.
Indeed,
$\la \alpha(H_{M_t}) | \alpha(H_{M_{t+1}}) \ra = \la \pi_t | \pi_{t+1} \ra
= \sum_{i} \sqrt{\pi_t(i) \pi_{t+1}(i)} \ge 1-|\pi_t-\pi_{t+1}|$
and therefore is non-negligible.
\end{proof}

Essentially all Markov chains that are used in approximate counting
that we are aware of meet the criteria of the theorem.
The following is a partial list of states we can Qsample from using
Theorem \ref{th3}, where the citations refer to
 the approximate algorithms that we use as the basis for the
quantum sampling algorithm:
\begin{enumerate}
\item Uniform
superposition over all perfect matchings of a given bipartite graph \cite{perm}.
\item All spanning trees of a given graph \cite{bubley}.
\item All lattice points contained in a high dimensional convex body
satisfying the conditions of \cite{applegatekannan}.
\item Various Gibbs distribution over rapidly mixing Markov chains using
the Metropolis filter \cite{lovasz}.
\item Log-concave distributions \cite{applegatekannan}.
\end{enumerate}

We note that most if not all of these states can be generated using other
simpler techniques.
However our techniques do not rely on self reducibility,
and are thus qualitatively different and perhaps extendible in other ways.
We illustrate our technique with the example of how to
Qsample from all perfect matchings in a given bipartite graph.
We also note that if we could relax the second requirement in Definition \ref{strong}
the techniques in this section
could have been used to give a quantum algorithm for  graph isomorphism.

\subsection{Qsampling from Perfect Matchings}
In this subsection we heavily rely on the work of Sinclair, Jerrum
and Vigoda \cite{perm} who recently showed how to efficiently approximate a
permanent of any matrix
 with non negative entries, using a sequence of Markov
chains on the set of
Matchings of a bipartite graph. The details of this work are far too
involved to explain here fully,
and we refer the interested reader to \cite{perm}
for further details.

In a nutshell, the idea in \cite{perm} is to apply a Metropolis random walk
 on the set of perfect and near perfect matchings (i.e. perfect matchings minus one edge) of the complete
bipartite graph. Since \cite{perm} is interested in a given input
bipartite graph, which is a subgraph of the complete graph,
 they assign weights to the edges such that edges
 that do not participate in the
input graph are slowly decreasing until the probability they appear in
the final distribution practically vanishes.
The weights of the edges are updated using data that is collected from
running the Markov chain with the previous set of weights, in an
adaptive way. The final Markov chain with the final parameters
converges to a probability distribution which is essentially
concentrated on the perfect and near perfect matchings of the input
graph, where the probability of the perfect matchings is $1/n$
times that of the near perfect matching.

It is easy to check that the Markov chains being used in \cite{perm}
are all strongly samplable, since they are Metropolis chains. Moreover,
the sequence of Markov chains is slowly varying.
It remains to see that can quantum sample from the limiting distribution of
the initial chain that is used in \cite{perm}.
This limiting distribution is a distribution over
all perfect and near perfect matchings
in the complete bipartite graph, with each near
perfect matching having weight $n$ times
 that of a perfect matching, where $n$ is the number of nodes
of the given graph.
Indeed, to generate this super-position we do the following:
\begin{itemize}
\item
We generate $\sum_{\pi \in S_n} \ket{m_\pi}$,
where $m$ in the matching on the bipartite graph induced by $\pi \in S_n$.
We can efficiently generate this state because we can generate a super-position over all permutations in $S_n$,
and there is an easy computation from a permutation to a perfect matching in a complete bipartite graph
and vice versa.
\item
We generate the state
$|0\ra+ \sqrt{n}\sum_{i=1}^n|i\ra$ normalized, on a $log(n)$ dimensional register.
This can be done efficiently because of the low-dimension.
\item
We apply a transformation
that maps $\ket{m,i}$ to
$\ket{0,m}$ when $i=0$, and to
$\ket{0,m-\{e_i\}}$ for $i>0$, where
$m-\{e_i\}$ is the matching $m$ minus the $i'th$ edge in the matching.
There is an easy computation from $m-\set{e_i}$ to $m,i$ and vice versa,
and so this transformation can be done efficiently.
We are now at the desired state.
\end{itemize}

Thus
we can apply Theorem \ref{th3} to Qsample from
the limiting distribution of the final Markov chain.
We then measure to see whether the matching is perfect or not,
and with non negligible probability we project the state onto
the uniform distribution over all perfect matchings of the given graph.

\section{Acknowledgements}
We wish to thanks Umesh Vazirani, Wim van Dam, Zeph Landau,
Oded Regev, Dave Bacon, Manny Knill, Eddie Farhi,
Ashwin Nayak and John Watrous
for many inspiring discussions.
In particular we thank Dave Bacon for an illuminating discussion
which led to the proof of claim \ref{sim}.

\appendix

\section{\bf Zeno effect approach to simulating adiabatic Generators}
\label{sec:zeno}

\begin{proof}(Of Claim \ref{cl})
We concentrate on a time interval $[s_0,s_1]$, $s_0 < s_1$, 
where $H(\cdot)$ is continuous on $[s_0,s_1]$ and differentiable
on $(s_0,s_1)$. We denote  $\eta_{max}=\max_{s \in (s_0,s_1)} || {dH \over ds}(s)||$
and $\Delta_{min} = \min_{s \in (s_0,s_1)} \Delta(H(s))$.
We choose $R \ge \Theta({\eta_{max}^2 \over \Delta_{min}^2} \epsilon)$.
Notice that $R$ is polynomially related to the schedule time $T$ in 
the adiabatic condition.

We divide the interval $[0,1]$ to $R$ time steps.
At time step $j$, $j=1,\ldots,R$, 
we measure the state with a projective, orthogonal measurement 
that has the ground state of $H({j \over R})$ as one of its answers.
We begin at the state $\alpha(H(0))$.

We need to show our procedure can be implemented efficiently,
i.e., that if $H$ is simulatable and has a non negligible spectral
gap, then such a measurement can be implemented efficiently.
We also need to show our procedure is accurate, i.e., 
that under the condition of the adiabatic theorem, for the $R$
we have chosen,  with very high probability the final state 
is indeed $\alpha(H(1))$.

\begin{description}
\item [Accuracy]:

We first bound the relative change of $H(s+\delta)$ with respect to $H(s)$.
For $s,s+\delta \in [s_0,s_1]$, $H(s+\delta)-H(s) = \int_{s}^{s+\delta} {dH \over ds}(s) ds$
and so $||H(s+\delta)-H(s)|| = ||\int_{s}^{s+\delta} {dH \over ds}(s) ds|| \le 
\int_{s}^{s+\delta} ||{dH \over ds}(s)|| ds \le \eta_{max} \cdot \delta$.

Our next step is to claim that two Hamiltonians that are close to each other have
close groundstates. This is captured in the following claim, that we prove later.

\begin{claim}\label{zen}
Let $H,J$ be two Hamiltonians $\|H-J\|\le \eta$.
Assume $H,J$ have large spectral gaps:  $\Delta(H),\Delta(J)\ge \Delta$
Then $|\la\alpha(H)|\alpha(J)\ra|\ge 1-\frac{4\eta^2}{\Delta^2}.$
\end{claim}

Having that, we see that since  $\|H({j+1 \over R})-H({j \over R})\|\le {\eta_{max} \over R}$,
Claim \ref{zen} asserts that the probability for
successful projection at the $j'th$
measurement, i.e. the probability that the outcome is indeed the groundstate,
is
$1-O({\eta_{max}^2 \over R^2 \Delta_{min}^2})$.
The probability we err at any of the $R$ steps is therefore at most
$O({\eta_{max}^2 \over R \Delta_{min}^2})$ which is at most $\epsilon$ by our choice of $R$. 

\item [Efficiency]:

We use Kitaev's phase estimation algorithm
\cite{kitaevphase,nielsen} to give, with polynomially good confidence, a polynomially good
approximation of the eigenvalue, and we then measure the eigenvalue.
As the spectral gap is non-negligible, this in effect does an
orthonormal measurement with the eigenstate subspace as one possible answer,
as desired.
The procedure is polynomial because $H$ is simulatable and
we can efficiently approximate $e^{-iHt}$ for every polynomial $t$.
\end{description}

\end{proof}

We finish with the proof of Claim \ref{zen}.

\begin{proof}(Of Claim \ref{zen})
W.l.o.g we can assume $H$ and $J$ are positive, 
otherwise just add $C \cdot I$ to both matrices,
for large enough constant $C$.
This does not effect the spectral norm of the difference, the spectral gaps or the inner product between
the groundstates.

Let $\set{v_i}$ be the $H$ eigenvectors with eigenvalues $\lambda_1 < \ldots < \lambda_N$,
and $\set{u_i}$, $\set{\mu_i}$ for $J$.
Again, w.l.o.g, $0=\lambda_1 \le \mu_1$.
Notice also that $\mu_1 \le \lambda_1+\eta$, because
$\mu_1=\min_{v:||v||=1} |Jv|$, and  $|J v_1| \le |H v_1|+|(J-H)v_1| \le \lambda_1+\eta$.

So, $|J v_1| \le \lambda_1+\eta$.
On the other hand, express $v_1=a u_1+ b u_{\bot}$,
with $u_{\bot} \bot u_1$.
Then, $|J v_1| = |b J u_\bot + a J u_1| ~\ge~
|b| \cdot \mu_2-|a| \cdot \mu_1 ~\ge~
|b| \cdot (\mu_1+\Delta)-|a| \cdot \mu_1 ~\ge~
|b| \cdot (\lambda_1+\Delta)-|a| \cdot (\lambda_1 + \eta)$.
Setting $\lambda_1=0$ we get:
$\eta \ge |b| \cdot \Delta - |a| \cdot \eta$.
Let us denote $c={\Delta \over \eta}$.
We see that $|a| \ge c |b|-1$.

We now plug in $|a|=\sqrt{1-|b|^2}$, and square both sides of the inequality.
We get
$1-|b|^2 \ge 1-2c|b|+c^2 |b|^2$, i.e., $|b| \le {2c \over c^2+1} \le {2c \over c^2}={2 \over c}$.
Equivalently, $| \la \alpha(H_1) | \alpha(H_2) \ra | = |\la v_1|u_1\ra|=|a| =
 \sqrt{1-|b|^2} \ge 1-{4 \over c^2}
= 1-{4 \eta^2 \over \Delta^2}$ as desired.
\end{proof}


\begin{thebibliography}{99}
\bibitem{aaronson} S. Aaronson, Quantum lower bound for the collision problem. STOC 2002, pp. 635-642
\bibitem{wim} D. Aharonov, W. van Dam, Z. Landau, S. Lloyd,
J. Kempe and O. Regev, Universality of
Adiabatic quantum computation, in preparation, 2002
\bibitem{aharonov2} D. Aharonov, A simple proof that Toffoli and Hadamard
are quantum universal, in preparation, 2002
\bibitem{mixing} N. Alon, Eigenvalues and Expanders. Combinatorica 6(1986), pp. 83-96.
\bibitem{alon} N. Alon and J. Spencer, The Probabilistic Method. John Wiley $\&$ Sons, 1991.
\bibitem{applegatekannan} D. Applegate and R. Kannan, Sampling and integration of near log-concave functions, 
STOC 1991, pp. 156--163
\bibitem{avron} Y. Avron and A. Elgart, Adiabatic Theorem without a Gap
Condition, {\it Commun. Math. Phys.} 203, pp. 445-463, 1999
\bibitem{born} M. Born, V. Fock and B. des Adiabatensatzes. Z. Phys. 51, pp. 165-169, 1928
\bibitem{bubley}
 R. Bubley and M. Dyer, Faster random generation of linear extensions, 
SODA 1998, pp. 350-354.
\bibitem{childs} A. M. Childs, R. Cleve, E. Deotto, E. Farhi, S. Gutmann and D. A. Spielman,
 Exponential algorithmic speedup by quantum walk, quant-ph/0209131
\bibitem{farhimeas}
    A. M. Childs, E. Deotto, E. Farhi, J. Goldstone, S. Gutmann, 
A. J. Landahl, Quantum search by measurement, 
      Phys. Rev. A 66, 032314 (2002)
\bibitem{farhi3}
  A. M. Childs, E. Farhi, J. Goldstone and S. Gutmann, Finding cliques by
  quantum adiabatic evolution, quant-ph/0012104.
\bibitem{walks}  A. M. Childs, E. Farhi and S. Gutmann,
An example of the difference between quantum and classical random walks,
quant-ph/0103020. Also, E. Farhi and S. Gutmann,  Quantum Computation and Decision Trees, quant-ph/9706062
\bibitem{legendre} W. van Dam and S. Hallgren, Efficient quantum algorithms for Shifted Quadratic Character Problems,
     quant-ph/0011067 
\bibitem{vandamvaz} W. van Dam, M. Mosca and U. V. Vazirani, How Powerful is Adiabatic Quantum Computation?. FOCS
       2001, pp 279-287
\bibitem{vandam2} W. van Dam and U. Vazirani, More on the power of adiabatic
computation, unpublished, 2001
\bibitem{farhi2}
  E. Farhi, J. Goldstone and S. Gutmann, A numerical study of the
  performance of a quantum adiabatic evolution algorithm for
  satisfiability, quant-ph/0007071.
\bibitem{farhi4}
  E. Farhi, J. Goldstone, S. Gutmann, J. Lapan, A. Lundgren and D. Preda,
  A quantum adiabatic evolution algorithm applied to random instances of
  an NP-complete problem, Science {\bf 292}, 472 (2001),
  quant-ph/0104129.
\bibitem{farhipaths}  E. Farhi, J. Goldstone and S. Gutmann,
Quantum Adiabatic Evolution Algorithms with Different Paths,
quant-ph/0208135
\bibitem{farhiad}  E. Farhi, J. Goldstone, S. Gutmann and
M. Sipser, Quantum Computation by Adiabatic Evolution,
quant-ph/0001106
\bibitem{GK93}O. Goldreich and E. Kushilevitz,
 A Perfect Zero-Knowledge Proof System for a Problem
                 Equivalent to the Discrete Logarithm,
Journal of Cryptology, pp. 97--116, vol 6 number 2, 1993
\bibitem{GG98} O. Goldreich and S. Goldwasser,
 On the limits of non-approximability of lattice
                 problems, STOC 1998, pp. 1-9
\bibitem{GSV98} O. Goldreich, A. Sahai and S. Vadhan,
Honest-Verifier Statistical Zero-Knowledge Equals
                 General Statistical Zero-Knowledge, STOC 1998 pp. 399--408,
\bibitem{GMR89}S. Goldwasser, S. Micali and C.
                 Rackoff, The Knowledge Complexity of Interactive Proof
                 Systems, SIAM Journal on Computing,
            SIAM J Comput, Vol 18 Number 1,  pp. 186--208, 1989
\bibitem{metro}M. Grotschel and L. Lovasz,
 Combinatorial Optimization: A Survey,
Handbook of Combinatorics, North-Holland, 1993
\bibitem{groversearch} L. Grover, Quantum Mechanics helps in searching for a needle in a haystack, 
Phys. Rev. Letters, July 14, 1997
\bibitem{grover} L. Grover and T. Rudolph, Creating superpositions that correspond to efficiently 
integrable probability distributions, quant-ph/0208112
\bibitem{hallgren}S. Hallgren, Polynomial-Time Quantum Algorithms for Pell's Equation and 
the Principal Ideal Problem, STOC 2002
\bibitem{perm} M. Jerrum and A. Sinclair, E. Vigoda
A Polynomial-Time Approximation Algorithm for the permanent of a matrix with non-negative entries, 
STOC 2000 
\bibitem{approx} M. Jerrum and A. Sinclair, The Markov chain Monte Carlo method: an approach to 
approximate counting and
     integration. in Approximation Algorithms for NP-hard Problems,
 D.S.Hochbaum ed., PWS Publishing, Boston,
     1996.
\bibitem{kato} T. Kato, On the adiabatic theorem of Quantum Mechanics,
J. Phys. Soc. Jap. {\bf 5}, pp. 435-439 (1951)
\bibitem{kitaevphase}  A. Yu. Kitaev,
Quantum measurements and the Abelian Stabilizer Problem,
quant-ph/9511026
\bibitem{gibook} J. Kobler, U. Schoning and J. Turan, The Graph Isomorphism Problem. Birkjauser, 1993.
\bibitem{landau} Landau and Lifshitz, {\it Quantum Mechanics}
 (Second edition of English Translation), Pergamon press, 1965
\bibitem{lovasz} L. Lovasz: Random Walks on Graphs: A Survey.
Combinatorics, Paul Erdos is Eighty, Vol. 2 (ed. D. Miklos, V. T.
     Sos, T. Szonyi, Janos Bolyai Mathematical Society,
Budapest, 1996, pp. 353--398.
\bibitem{masia} Messiah, {\it Quantum Mechanics}, John Willey $\&$ Sons (1958)
\bibitem{nielsen}M. A. Nielsen and I. Chuang,
                                    Quantum Computation and Information,
                                   Cambridge
                                    University Press, 2000
\bibitem{zeno}  See A. Peres, Quantum Theory: Concepts and methods, Kluwer Academic Publishers, 1995
\bibitem{cerf}J. Roland and N. Cerf, Quantum Search by Local Adiabatic Evolution
      Phys. Rev. A 65, 042308 (2002)
\bibitem{SV97} A. Sahai and S. P. Vadhan,
A Complete Promise Problem for Statistical
                 Zero-Knowledge, FOCS 1997 pp. 448--457
\bibitem{shor} P. W. Shor: Polynomial-Time Algorithms for Prime Factorization and Discrete Logarithms on a Quantum
        Computer. SIAM J. Comput. 26(5) 1997, pp. 1484-1509
\bibitem{spitzer} W. L. Spitzer and S. Starr,
Improved bounds on the spectral gap above frustration free ground states of
quantum spin chains, math-ph/0212029
\bibitem{shi} Y. Shi, Both Toffoli and Controlled-NOT need little help to do universal quantum
computation, quant-ph/0205115
\bibitem{V00} S. Vadhan, A Study of Statistical Zero Knowledge Proofs,
PhD Thesis, M.I.T., 1999
\bibitem{watrous} J. Watrous: Quantum algorithms for solvable groups. STOC 2001, pp. 60-67
\end{thebibliography}
\end{document}